\pdfoutput=1
\documentclass[iop]{emulateapj}
\usepackage{graphicx,epsfig}
\usepackage{rotate}
\usepackage{multirow}
\usepackage{morefloats}
\newcommand{\bq}{\begin{equation}}
\newcommand{\eq}{\end{equation}}

\def\gtsim{\lower.5ex\hbox{$\buildrel > \over\sim$}}
\def\ltsim{\lower.5ex\hbox{$\buildrel < \over\sim$}}

\def\sun{\mbox{$_\odot$}}
\def\apjl{ApJL}
\def\apj{ApJ}
\def\apjs{ApJS}
\def\mnras{MNRAS}
\def\araa{ARAA}

\def\aap{A\&A}
\def\aaps{A\&A Suppl.}
\def\nat{Nature}

\shorttitle{PISN Rotation}
\shortauthors{Chatzopoulos, Wheeler, Couch}
\begin{document}
\title
{MULTIDIMENSIONAL SIMULATIONS OF ROTATING PAIR-INSTABILITY SUPERNOVAE}
\author{E. Chatzopoulos\altaffilmark{1}, J. Craig Wheeler\altaffilmark{1} \& Sean M. Couch\altaffilmark{2,3}}
\altaffiltext{1}{Department of Astronomy, University of Texas at Austin, Austin, TX, 78712, USA.}
\altaffiltext{2}{Department of Astronomy \& Astrophysics, Flash Center for Computational
Science, University of Chicago, Chicago, IL, 60637, USA.}
\altaffiltext{3}{Hubble Fellow}

\begin{abstract}
We study the effects of rotation on the dynamics, energetics and $^{56}$Ni production of Pair Instability
Supernova explosions by performing rotating two-dimensional (``2.5-D") hydrodynamics simulations. 
We calculate the evolution of 
eight low metallicity ($Z =$~10$^{-3}$, 10$^{-4}$~$Z_{\odot}$) massive (135-245~$M_{\odot}$) PISN progenitors
with initial surface rotational velocities 50\% that of the critical Keplerian value using the stellar evolution code {\it MESA}. 
We allow for both the inclusion and the omission of the effects of magnetic fields in the 
angular momentum transport and in chemical mixing, resulting in slowly-rotating and rapidly-rotating final 
carbon-oxygen cores, respectively. 
Increased rotation for carbon-oxygen cores of the same mass and chemical stratification leads to less 
energetic PISN explosions that produce smaller amounts of $^{56}$Ni due to the effect of the angular momentum
barrier that develops and slows the dynamical collapse. We find a non-monotonic dependence of $^{56}$Ni production
on rotational velocity in situations when smoother composition gradients form at the outer edge of the rotating cores.
In these cases, the PISN energetics are determined by the competition of two factors: the extent of chemical mixing in the
outer layers of the core due to the effects of rotation in the progenitor evolution and the development of
angular momentum support against collapse. 
Our 2.5-D PISN simulations with rotation are the first presented in the literature. They reveal 
hydrodynamic instabilities in several regions of the exploding
star and increased explosion asymmetries with higher core rotational velocity.
\end{abstract}


\vskip 0.57 in

\section{INTRODUCTION}\label{intro}

Pair Instability Supernovae (PISNe) are triggered by the development of a dynamical instability
in the carbon-oxygen (CO) cores of massive stars that enter a regime of high temperature ($\sim$~$10^{9}$~K) and relatively
low density ($10^{3}$-$10^{6}$~g~cm$^{-3}$) 
favoring the substantial production of electron-positron (e$^{+}$e$^{-}$) pairs. When the density of electron-positron
pairs becomes high the volume averaged adiabatic index decreases ($\Gamma_{1}<$~4/3) eventually triggering dynamical collapse followed by
thermonuclear burning of C and O. This subsonic burning produces enough energy to either totally disrupt 
the progenitor star (full-fledged PISN) or remove the outermost, less gravitationally bound layers in the form
of a pulsational PISN (PPISN; Woosley, Blinnikov \& Heger 2007) depending primarily on the mass of the final CO core, $M_{\rm CO}$.
In addition, large amounts of radioactive $^{56}$Ni ($\sim$~1-60~$M_{\odot}$; Heger \& Woosley 2002) are produced that
power the resulting SN light curve (LC).

The hypothesis that PISNe are the ultimate fate of very massive
stars ($M_{ZAMS}>$~80~$M_{\odot}$; ZAMS: Zero Age Main Sequence) was introduced more than a half century ago
(Rakavy \& Shaviv 1967; Barkat, Rakavy \& Sack 1967; Rakavy, Shaviv \& Zinamon 1967, Fraley 1968).
Stellar evolution models of non-rotating PISN progenitors and spherical hydrodynamics simulations 
of these explosions have been presented in various contexts ever since 
(Ober, El Eid \& Fricke 1983, Fryer, Woosley \& Heger 2001; Heger \& Woosley 2002). 
Numerical LCs and spectra for PISNe
have also been presented, based on one-dimensional simulations 
(Scannapieco et al. 2005, Kasen, Woosley \& Heger 2011; Whalen et al. 2012; Dessart et al. 2013).
Recently, a few preliminary 2-D simulations of non-rotating PISNe have been presented that investigate
the effects of mixing due Rayleigh-Taylor (RT) instabilities that develop between the CO
core and the surrounding He layer of the star and also in the outer regions due to the reverse shock 
that develops after the SN blast wave exits the stellar surface (Joggerst \& Whalen 2011; 
Chen, Heger \& Almgren 2011; 2012).

Observational evidence suggests the existence of very massive stars (VMS; $M_{ZAMS} <$
~320~$M_{\odot}$)
capable of producing PISN (Crowther et al. 2010). In addition, PISNe are thought to be related to some
superluminous supernovae (SLSNe; Gal-Yam 2012 and references therein), especially SN~2007bi (Gal-Yam et al. 2009).
The nature of SN~2007bi is a topic still under debate centered around the radiative properties of PISNe 
(Chatzopoulos \& Wheeler 2012b; Dessart et al. 2013). 
PPISN have also been discussed in the context of SLSNe, where the exceptional
luminosity is produced by the interaction between multiply ejected PPISN shells 
(Woosley, Blinnikov \& Heger 2007; Chatzopoulos \& Wheeler 2012b). 

The nucleosynthetic output and energetics of PISNe are of great importance to early Universe studies 
where a significant number of Population III stars are found to have masses in the PISN-producing regime in some models
(Abel et al. 1998; Abel et al. 2000; Bromm et al. 2002; Bromm \& Larson 2004). 
Future missions like the James Webb Space Telescope ({\it JWST}), {\it WFIRST} and the Wide-Field Infrared
Surveyor for High-Redshift ({\it WISH}) are set to look for these
primordial PISN explosions that are responsible for enriching the primordial interstellar medium
(Scannapieco et al. 2005; Pan, Kasen \& Loeb 2012, Hummel et al. 2012, Whalen et al. 2012, 2013).
The number of Population III stars that produce PISNe may be reduced if fragmentation
is important  (Stacy et al. 2010; Greif et al. 2011), but rapid pre-PISN rotation may counter this effect allowing
for the production of massive CO cores from lower ZAMS mass stars (Chatzopoulos \& Wheeler 2012a, Yoon,
Dierks \& Langer 2012). Rapid rotation is found to be present in star formation simulations of Population III
stars (Greif et al. 2011; Stacy et al. 2013).

Woosley, Blinnikov \& Heger (2007) estimated the minimum ZAMS mass limits for
PPISN to be 95~$M_{\odot}$~$<M_{ZAMS}<$~130~$M_{\odot}$ and for PISN to be
130~$M_{\odot}$~$<M_{ZAMS}<$~260~$M_{\odot}$ in the case of zero rotation
and solar metallicity but with an ad-hoc pre-PISN mass loss assumption. 
Heger \& Woosley (2002) present these mass limits in terms of the final $M_{\rm CO}$
(PPISN: 40~$M_{\odot}$~$<M_{CO}<$~60~$M_{\odot}$,
PISN:  60~$M_{\odot}$~$<M_{CO}<$~137~$M_{\odot}$).
Langer et al. (2007) estimate there are no massive stars with $Z>Z_{\odot}/3$
capable of producing PISNe because the effects of the extreme mass-loss they
experience during their evolution. 

The ZAMS mass limits for PISNe are reduced if the effects of rotation are taken into
account in the progenitor evolution (Chatzopoulos \& Wheeler 2012a; Yoon, Dierks \&
Langer 2012; Yusof et al. 2013). Rotational mixing, mainly due to meridional circulation
but also the effects of the magnetic field, if considered
(Spruit-Tayler dynamo; Spruit 1999, 2002 hereafter ``ST"), can recycle unprocessed
material from the progenitor's outer envelope into the core, thus increasing the fuel available
for nuclear reactions. Rotation in massive stars is found to lead to a bluer, more luminous
evolution in the Hertzsprung-Russell (HR) diagram and
to chemically homogeneous evolution (CHE) also in the context of
core-collapse supernovae (CCSNe) and Gamma-ray burst (GRB) progenitors
(Heger, Langer \& Woosley 2000; Yoon \& Langer 2005; Heger, Woosley \& Spruit 2005; Woosley \& Heger 2006; 
Esktr{\"o}m et al.  2008; Maeder \& Meynet 2011; Brott et al. 2011a; Brott et al. 2011b; Esktr{\"o}m et al.  2008).
The CHE has the effect of allowing the production of massive final CO cores from 
less massive ZAMS progenitors than those required in the case of no rotation. 
Chatzopoulos \& Wheeler and Yoon, Dierks \& Langer derived the mass limits
for PPISNe and PISNe in the case of rotating Population III stars to be
$\sim$~50~$M_{\odot}$~$<M_{ZAMS}<$~85~$M_{\odot}$ for PPISN
and 
$\sim$~85~$M_{\odot}$~$<M_{ZAMS}<$~190~$M_{\odot}$ for PISN
progenitors rotating at 50\% the critical Keplerian rate, $\Omega_{c}$,
where $\Omega_{c} = (g(1-\Gamma_{Ed})/R)^{1/2}$ with
$g =$~$GM/R^{2}$ the gravitational 
acceleration at the ``surface" of the star, $G$ the universal gravitational constant, $M$ the
mass, $R$ the radius of the star and $\Gamma_{Ed}=L/L_{Ed}$ the Eddington factor where 
$L$ and $L_{Ed}$ are the total radiated luminosity and the Eddington luminosity, respectively.
The mechanical effects of the centrifugal force during dynamical collapse can also effect
the PISN mass limits: increased rotation can decelerate collapse and even lead to escape
a full-fledged PISN explosion, in the most extreme cases (Glatzel et al. 1985). This has as a result the shift of the PISNe regime to higher
mass limits, however the magnitude of this effect is not established.
Rapid rotation ($\Omega/\Omega_{c} \geq$~50\%) is observed in nearby massive stars 
(Dufton et al. 2011).

The effects of rotation on the dynamics of the explosions themselves (the dynamical collapse,
the hydrodynamical instabilities that develop, as well as the energetics and nucleosynthetic
signature) have not yet been explored in detail, and only a few efforts have been
presented in the past in the case of spherical geometry 
(Glatzel et al. 1985; Stringfellow \& Woosley 1988). Glatzel et al. explored rotation in PISNe using
the method of Maclaurin spheroids and found that rigid-body rotation leads to more oblate explosions and less complete explosive
oxygen burning. For high degrees of rotation the collapse does not lead to explosion by means of a full-fledged PISN.
In the present work, we have used the stellar evolution code {\it MESA}
(Paxton et al. 2011; 2013) to evolve massive rotating PISN progenitor stars and the new version of the Adaptive
Mesh Refinement (AMR) hydrodynamics code {\it FLASH} (Fryxell et al. 2000),
that includes a treatment for
rotation, to study the explosion properties of rotating PISNe in ``2.5-D" (a two-dimensional
grid plus the rotational velocity vectors in the perpendicular direction).

The paper is organized as follows: a presentation of the {\it MESA} pre-PISN
progenitor evolution calculations is given in \S2, the 2.5-D {\it FLASH} hydrodynamics
PISN simulations and results are presented in \S3 and finally our discussions and conclusions
are summarized in \S4.

\section{PRE-PISN EVOLUTION WITH {\it MESA}}\label{progevol}

In order to produce physically consistent rotating PISN progenitors for the 2.5-D hydrodynamics
simulations the first step is to evolve a grid of massive stars that produce a variety of final
core rotation velocities, $v_{\rm rot,c}$. 
The stellar evolution calculations were done using the modular code {\it MESA} version 4631
(Paxton et al. 2011; 2013). In all of our {\it MESA} calculations the standard mass-loss rate
prescriptions appropriate for massive stars are used (de Jager et al. 1988; Vink et al. 2001). 
We use the Timmes \& Swesty (2000) ``Helmholtz" equation of state (HELM EOS) that includes the 
contributions from e$^{+}$e$^{-}$ pairs and the ``approx21" nuclear reaction network (Timmes 1999) that 
includes the $\alpha$-chain elements, and the intermediate elements linking those through $(\alpha, p)(p,\gamma)$
reactions from neutrons and protons all the way up to $^{56}$Ni (mass number $A$ from 1 through 56). Those
input assumptions are very similar to the ones used for the non-rotating {\it MESA} PISN progenitors
presented by Dessart et al. (2013). For the treatment of convection the Schwarzschild criterion is adopted
with the choice of $\alpha_{\rm MLT} =$~2 for the mixing length. 

Rotation in {\it MESA} is treated using the prescriptions of Heger, Langer \& Woosley (2000)
and Heger, Woosley \& Spruit (2005) that include many relevant hydrodynamical instabilities
that affect the mixing of chemical species and angular momentum transport (namely
the meridional circulation, the dynamical and secular shear instabilities and the 
Solberg-Hoiland and Goldreich-Schubert-Fricke instabilities). {\it MESA} has also the
capability of including the effects of magnetic fields on angular momentum transport
and mixing of species based on the ST prescriptions (Spruit 1999; 2002). The effects
of rotation on mass-loss are also treated using the approximation presented in Heger, Langer \& Woosley:
$\dot{M} = \dot{M}_{\rm no-rot}/(1-\Omega/\Omega_{c})^{0.43}$ where $\dot{M}_{\rm no-rot}$ is the mass-loss
rate in the case of no rotation and $\Omega$ is the surface angular velocity at the stellar equator. 
For cases approaching $\Omega/\Omega_{c} =$~1 the mass loss calculated using the above formula
diverges. For this reason in {\it MESA}, following Yoon et al. (2010), the mass loss timescale is limited
to the thermal timescale of the star, $\tau_{KH}$: $\dot{M} = min[\dot{M}(\Omega),f M/\tau_{KH}]$ where
$f$ is an efficiency factor taken to be 0.3.

In order to better probe the effects of rotation and to include a variety of progenitor characteristics we
ran eight models that span two metallicity series ($Z =$~10$^{-3}$~$Z_{\odot}$ and
$Z =$~10$^{-4}$~$Z_{\odot}$) with four models run for each metallicity. The evolution of all models was initiated 
in the pre-main sequence phase (pre-MS) without rotation up to ZAMS.  Then the desired degree of rotation (50\% the critical 
value in all cases) was introduced at the ZAMS for the remainder of the evolution. 
The evolution was stopped for all models at the same stage of nuclear burning
upon encountering the e$^{+}$e$^{-}$ pair instability and at a point where a significant mass fraction of the stellar cores
was within the $\Gamma_{1} <$4/3 regime in the density-temperature ($\rho$-$T$) plane. At this stage, {\it MESA} has the capability of
computing subsonic hydrodynamical effects and can follow the dynamical collapse up until central $^{20}$Ne exhaustion
($X_{Ne,c} =$~0.01 where $X_{Ne,c}$ is the neon mass fraction at the central zone of the model) and before $^{16}$O burning
is initiated, where we formally interrupt the evolution. This is the same criterion used in the PISN progenitor models presented
by Ober, El Eid \& Fricke (1983) and Dessart et al. (2013). Using this termination criterion, we are provided with a set of
PISN progenitor models at very similar nuclear burning and hydrodynamical stages that we subsequently map to the
AMR grid of {\it FLASH}. For all models a high degree of resolution is chosen resulting in final PISN progenitor models
with 4,000-6,000 grid points (the ``mesh\_delta\_coeff" in {\it MESA} was given values 0.35-0.75). This reasonable resolution
was necessary in order to properly resolve convection and burning processes 
during the advanced burning stages months to days before the onset of the dynamical instability. 

For each metallicity series we ran four different models:
(a) a non-rotating model (``norot"), 
(b) a model with 50\% critical rotation at the ZAMS and the ST effects
for the magnetic field included (``rotST"), 
(c) a model with 50\% critical rotation at the ZAMS with the ST effects
omitted (``rotnoST") and 
(d) a model with 50\% critical rotation at the ZAMS, the ST effects included and with a higher
adopted mass-loss rate parameter used (``rotST\_ml2").
The reason that we chose to run models both for the rotST and the rotnoST cases is because
we aim to obtain final CO cores of the same mass but with different rotational profiles. This allows us
to study the effects of different degrees of rotation that result from a self-consistent evolution
process. The inclusion of magnetic fields (``rotST" models) imposes magnetic torques and
magnetic viscosity that can significantly slow down the rotation of the core. The slow core rotation
predicted by the ST treatment seems to be consistent with the observed rotation rates
of some low mass stars, but also isolated white dwarfs, 
suggesting that the cores of some massive stars undergo an angular momentum loss process
prior to the explosion (Kawaler 1988; Heger, Woosley \& Spruit 2005; Suijs et al. 2008). 
Magnetic braking (Meynet, Eggenberger \& Maeder 2011) 
has been suggested as an alternative
explanation for this observation. No direct evidence suggests that 
the same holds for the cores of VMSs and progenitors of PISNe that experience a much different
fate than CCSNe and never reach the burning stages all the way up to Fe. 
In addition, the efficiency of the ST dynamo mechanism is still much debated (Zahn et al. 2007).
This allows us to consider the case where there is no core spin-down process (``rotnoST") and
rapidly-rotating CO cores are formed. 

We also run ``rotST\_ml2" models with higher adopted mass-loss rate parameter
to investigate the competing effects of mass loss and rotationally-induced mixing. 
Increased mass-loss during the pre-PISN evolution reduces the smoothing of composition gradients
due to rotationally-induced mixing at the interface between the CO core
and the outer He envelope. This leads to differences in the dynamical collapse as we discuss 
in \S3 in detail. For the hydrodynamical analysis, we chose to include
models that do not include the effects of ST, but for which the rotational velocities were
artificially set to zero prior mapping to {\it FLASH} (``rotnoST\_v0" models) 
in order to study the effects of rotation in otherwise structurally identical PISN progenitors.

For each metallicity series, the masses of the four models were chosen carefully
after a number of trials, so that the resulting final CO cores had almost identical $M_{\rm CO}$.
The evolution of all models spanned the range 2.5-3.5~Myr.
In the case of $Z =$~10$^{-3}$~$Z_{\odot}$ models 200sm\_norot, 140sm\_rotST,
135sm\_rotnoST and 150sm\_rotST\_ml2, with ZAMS masses 200, 140,
135, and 150~$M_{\odot}$ respectively, all produced final CO cores with
$M_{CO} \simeq$~80~$M_{\odot}$.
For $Z =$~10$^{-4}$~$Z_{\odot}$ models 245sm\_norot, 205sm\_rotST,
195sm\_rotnoST and 217sm\_rotST\_ml2, with ZAMS masses 245, 205,
195, and 217~$M_{\odot}$ respectively, all produced final CO cores with
$M_{CO} \simeq$~100~$M_{\odot}$. Therefore all final
CO cores have masses deep within the regime predicted for full-fledged
PISN explosions. The basic physical characteristics of all {\it MESA} models
presented here are given in Table 1
($M_{\rm ZAMS}$, the final pre-explosion mass $M_{f}$, $\Omega/\Omega_{c,X,Y}$ where $X =s,c$ for ``surface" and
``core edge", respectively and $Y =i,f$ for the initial (ZAMS) and final
pre-PISN values, $v_{\rm rot,c}$ and total stellar binding energy of the final model, $E_{b,f}$ 
and radius $R_{f}$. In all cases the ``edge" of the CO core is defined as the point where
the sum of the $^{12}$C and $^{16}$O mass fractions drop below 0.5, $X_{C} + X_{O} <$~0.5).

The basic structural characteristics of all PISN progenitor models at the time the {\it MESA} calculation was
terminated are plotted in Figures 1 through 6. In all figures black curves are for the ``norot," red curves
for the ``rotST," blue curves for the ``rotnoST" and green curves for the ``rotST\_ml2" models.
Note how the CO core $\rho$, $T$ and composition ($X_{i}$) structures are strikingly similar for all models of
the same metallicity group (the CO core radii, $R_{CO}$, range from $0.2-1 \times 10^{11}$~cm for all models).
Exceptions are the rotational (Figures 1,2,5) and radial (Figures 1,2) velocity profiles. 
The rotational velocity profiles show significantly higher rotational velocities for the cores of the
``rotnoST" models ($>$~1000~km~s$^{-1}$).
In retrospect, moderate to low rotational velocities are obtained for the ``rotST" and ``rotST\_ml2" models
($\sim$~60-200~km~s$^{-1}$) and the core-envelope coupling is broken.
The radial velocity profiles indicate that
at the time of mapping to {\it FLASH} the cores of all models are in the phase of dynamical collapse with infall velocities
exceeding 500~km~s$^{-1}$. Although differences of the order of $\sim$~100~km~s$^{-1}$ exist in the radial
velocity profiles between models of the same $M_{\rm CO}$, we discuss in \S3 that these differences do not alter the final
PISN energetics and $^{56}$Ni yields significantly. 

Another important thing to note in Figures 3-4 is the compositional gradient differences 
at the interfaces between the CO cores and the overlying He layers of all models
(more specifically at mass coordinates 75-85~$M_{\odot}$ for the $Z =$~10$^{-3}$~$Z_{\odot}$
and 95-105~$M_{\odot}$ for the $Z =$~10$^{-4}$~$Z_{\odot}$ series). This argument is better
illustrated in \S3.1 in terms of the structure of the mean molecular weight, $\mu$.
Non-rotating models produce progenitors with classic, well-defined onion-structures 
between layers with different composition and clear distinctions between the CO core and the H/He envelope. 
On the other hand, the core to envelope transition for ``rotST" and ``rotnoST" models generally has smoother
composition gradients (specifically for the He, C, N, O and Ne species). This behavior is reversed again
for the ``rotST\_ml2" models where the rapid mass-loss rate adopted in the evolution calculation
prevents effective rotational mixing. This leads to steeper composition gradients in the core-envelope
interfaces for the ``rotST\_ml2" models. 
Also, the core-envelope compositional transition for the ``rotST" models is smoother
in the 10$^{-3}$~$Z_{\odot}$ model series than in the 10$^{-4}$~$Z_{\odot}$ series indicating
that the uncertain effects of mass-loss for different degrees of metallicity can significantly
affect the ability of a rotating massive star to undergo efficient chemical mixing. 

\section{2.5-D {\it FLASH} SIMULATIONS}\label{sims}

The {\it MESA} PISN progenitor models discussed in \S2 were then mapped to the 2-D AMR grid
of the multi-physics hydrodynamics code {\it FLASH} (Fryxell et al. 2000; Dubey et al. 2009) in order to follow
the dynamical collapse, the nucleosynthetic production and possible hydrodynamic 
instabilities that develop. 
The latest release of {\it FLASH} is used (version 4.0) with the implementation of the
new unsplit PPM hydrodynamic (UHD; Lee \& Deane 2009) 
solver that allows for the inclusion of angular momentum. The basic physics units implemented
in {\it FLASH} and used for our simulations are nearly identical to the ones used in {\it MESA} 
(the ``Helmholtz" EOS and the ``Aprox19" nuclear reaction network). We also use the
new and updated Poisson multipole gravity solver. 

The mapping from the 1-D {\it MESA} Lagrangian grid in spherical coordinates to the 2-D {\it FLASH}
Eulerian grid in cylindrical coordinates is done carefully by first converting the cell-outer-edge values
for radius, $v$ and $v_{\rm rot}$ in the {\it MESA} outputs to cell-center averaged values. 
In each case the entire PISN progenitor star is included in the simulation box. A smooth $r^{-2}$ wind
with mass-loss rate $\dot{M}=$~$10^{-5}$~$M_{\odot}$~yr$^{-1}$ and wind velocity $v_{w} =$~100~km~s$^{-1}$ 
is joined to the outer edge of all PISN progenitors for the {\it FLASH} simulations.
We then select the proper resolution in {\it FLASH} that provides good conservation of total mass, energy and angular
momentum. For the entire duration of the simulations presented here all these global quantities were found to be
conserved at the $\sim$~10$^{-5}$ level 
with minor deviations attributable to flows outside of the simulation box and numerical error.  
Additionally, the grid resolution is chosen carefully in order to properly resolve
the $^{12}$C and $^{16}$O burning and subsequent flame propagation as well as hydrodynamical
instabilities. For this, a resolution study was done yielding scales resolved down to $10^{7}$-$10^{8}$~cm
for all models. This spatial resolution is very similar to that suggested by
Chen, Heger \& Almgren (2011; 2012). In our 2-D grids, this resolution corresponds
to a total of $10^{6}$-$10^{7}$ zones, depending on the model simulated.
We ran all of our {\it FLASH} simulations at the Texas Advanced Computing Center (TACC) {\it Stampede}
supercomputer using a total of $\sim$~10,000 CPU hours. Visualization of the simulation data 
was done using the {\it VisIt} version 2.6 software also run in parallel mode on the {\it Stampede}
supercomputer.

The rotational velocities computed from {\it MESA} were mapped to the 2-D grid of {\it FLASH} 
in cylindrical coordinates as
vectors with a direction perpendicular to the R-z plane of the simulation and the rotation axis
being coincident with the polar axis. To properly account for the treatment of shellular rotation in mapping
1-D rotational velocities from {\it MESA} into the 2-D grid of {\it FLASH}, the following formula was
used:

\begin{equation}
v_{\rm rot} (R,z) = \Omega(r) R,
\end{equation}
where $\Omega(r) = v_{\rm rot}(r)/r$,
$v_{\rm rot}(r)$ is the {\it MESA} 1-D rotational velocity and $r$ is the spherical
radial coordinate ($r=(x^{2}+y^{2})^{0.5}$). Recall that the angular velocity, 
$\Omega (r)$, is a constant for a particular spherical shell in the shellular approximation.
Formally, shellular rotation is defined as constant $\Omega$ on equipotential surfaces (Zahn 1992, Meynet \& Maeder 1997). 
In the cases we study here, however, the angular velocities in the cores of the model never exceed 40\% of the critical value
and the equipotential surfaces are close to spherical.
The result of this rotational velocity mapping scheme to the 2-D cylindrical-coordinate grid of {\it FLASH}
is shown in Figure 7 for the 140sm\_rotST model.

 The simulations were run until after the SN shock breaks out of the stellar surface in each model. 
 For all models, upon mapping to 
 {\it FLASH} the dynamical timescale to collapse was effectively the free-fall timescale 
 ($\sim$~50-100~s). 
 The $^{16}$O mass fraction and density structures at times of the order of $\sim$~100~s after
 dynamical collapse are presented in Figures 8-11 for the 10$^{-3}$~$Z_{\odot}$ model series
 and in Figures 12-15 for the 10$^{-4}$~$Z_{\odot}$ model series. The development of 
 RT instability between the oxygen core and the H/He envelope
 is clearly visible for all models with the exception of model 140sm\_rotST. The latter
 is due to the fact that the H/He envelope is very thin in this PISN progenitor, and as a result the
 interaction between the core and envelope material during collapse is weak, suppressing
 the development of prominent RT fingers. 
 
 The appearance of the RT instability throughout the
 PISN explosions results in mild chemical mixing between the core and envelope material
 and is in agreement with the findings of Chen, Heger \& Almgren (2011; 2012) and 
 Joggerst \& Whalen (2011). The RT mixing appears to be stronger at the CO core - H/He envelope
 interface than in the inner regions composed mostly of newly formed $^{56}$Ni. 
 The other aspect to note in Figures 8-15 is the significant explosion asymmetry with increased
 rotation. This can be seen in the left panels of Figures 10 and 14 where the $^{16}$O mass fraction
 maps for the ``rotnoST" models are presented. 
 The angular momentum barrier works on the collapse to keep the equatorial material
 from compressing as much, thus burning less, and hence afterwards expanding less rapidly. 
As a result, the inner core regions where $^{16}$O is exhausted and $^{56}$Ni has formed in its place 
take an oblate shape for these models. 

Figures 16-17 show the $^{56}$Ni mass fraction for all models at $\sim$~150~s after
maximum compression. The mild RT mixing between the newly formed $^{56}$Ni
and the remainder $^{16}$O-rich material as well as the increased asymmetry
with rotation discussed above are also distinguishable in these figures. 
The amplified explosion asymmetry with rotation can also be seen in Figures 18-19
where the density distributions are plotted for all models at the time the SN shock breaks
out of the stellar surface. The case of the fast rotator 135sm\_rotnoST stands out where
the SN blast wave breaks out from the polar regions at an earlier time than from the
equator. In addition, the formation of a reverse shock after the SN blast wave breaks out
of the stellar envelope leads to the formation of weak RT instability in the outer stellar
regions also in accordance to the effects noted by Chen, Heger \& Almgren (2011; 2012). 

Upon completion of all simulations we were able to
calculate the central density and central temperature
($\rho_{c}$ and $T_{c}$) hydrodynamic tracks, the total explosion energy and 
total mass of nucleosynthetic products, such as $^{56}$Ni, for all models.
Figure 20 shows the evolution of $\rho_{c}$ and $T_{c}$ upon mapping to {\it FLASH}.
At first inspection, the $\rho_{c}$-$T_{c}$ tracks look very similar among models of
the same metallicity series, however, a more detailed analysis reveals that
the central values reached in each case are sufficiently different to induce
differences in the burning rates, that are all
very sensitive to $T$. This is reflected by the final $^{56}$Ni mass produced
in each case that is plotted in Figure 21. 
The PISN explosion energies for all models varied in the range $10^{52}$-$10^{53}$~erg.
Table 2 lists some characteristics of the PISN explosions discussed here. 

As can be seen in Figure 21,
otherwise structurally identical models produce stronger PISNe with more $^{56}$Ni
when the rotational velocities are set to zero: 
``rotnoST\_v0" models produce 2-10 times more $^{56}$Ni
than ``rotnoST" models. This is simply because the rotational support from
the centrifugal force was artificially removed for the ``rotnoST\_v0" models prior to 
mapping to {\it FLASH}, thus leading to a stronger collapse. 
On the other hand, we derive a non-monotonic $^{56}$Ni production
as a function of CO core rotational velocity for the ``norot", ``rotST" and ``rotnoST"
models derived from different evolutionary patterns: adding a small rotation (``rotST" models)
seems to lead to more energetic PISNe that produce larger amounts of $^{56}$Ni, but
adding even more rotation leads to much weaker explosions. We discuss this
counter-intuitive result below in \S 3.1.
\subsection{{\it The source of the non-monotonic $^{56}$Ni production.}}

The non-monotonicity of $^{56}$Ni production for models of the same $M_{\rm CO}$
but different rotational profiles holds for both metallicity model series studied here and
is a puzzling result. Intuitively, adding rotation
to a collapsing star should lead to weaker PISNe because of the presence of an
angular momentum barrier that decelerates dynamical collapse, especially in the
equatorial regions. Therefore increasing rotation is expected to lead to smaller
$\rho_{c}$ and $T_{c}$ values at maximum compression, 
lower peak reaction rates and eventually less $^{56}$Ni produced.

To understand the source of the $^{56}$Ni non-monotonicity we focus on the
``norot" and ``rotST" models of both metallicity series. 
First we discuss the omission of the magnetic field in the mapping of ``rotST" models
in {\it FLASH}. Neglecting the B-field corresponds to assuming zero
magnetic pressure in the momentum equation that can in turn lead to overestimating the
inward acceleration. The {\it MESA} evolution allows for the calculation
of the radial ($B_{r}$) and toroidal ($B_{\phi}$) components of the magnetic
field in the star using the Spruit (1999, 2002) prescriptions. In the
cores of the ``rotST" models the total B-field values ($B = (B_{r}^{2}+B_{\phi}^{2})^{0.5}$)
are of the order of $10^{8}$~G, corresponding to magnetic pressures ($\sim$~$B^{2}/4\pi$)
of $\sim$~$10^{15}$~dyn~cm$^{-2}$. 
The total gas and radiation pressure in the same regions
are of the order of $\sim$~$10^{18}$-$10^{23}$~dyn~cm$^{-2}$, several orders of magnitude
higher, with corresponding plasma parameters $\beta \sim$~$10^{-3}$~$10^{-8}$,
indicating that the effects of magnetic pressure in the dynamics are negligible.

We then investigate whether multi-dimensional effects, such as large scale ``plume" mixing,
the development of hydrodynamical instabilities and directional effects
due to rotation can be the source of the non-monotonicity. A caveat
in this argument is that the dynamical time-scales upon mapping
to {\it FLASH} are likely shorter than relevant 2-D mixing time-scales
that might have an effect on the dynamics. Nevertheless we
test this hypothesis by re-running the ``norot" and ``rotST" models
in the 1-D spherical grid of {\it FLASH} with the rotational velocity
vector perpendicular to the radial coordinate (``1.5-D" treatment).
Note that in this case the mapping of the {\it MESA} $v_{\rm rot}$
values to the {\it FLASH} grid is straightforward, unlike in the
2-D cylindrical mapping scheme we described earlier. 
The 1.5-D simulations yielded final $^{56}$Ni masses
very close to those found in the 2.5-D treatment. Therefore
the non-monotonic $^{56}$Ni production trend
with rotational velocity remained unaltered. 

Having eliminated multi-dimensional effects and the
neglect of B-fields as the sources of the dynamical
collapse differences between the ``norot" and ``rotST" models,
we turn our attention to structural differences in the initial
{\it MESA} models. A careful inspection of Figures 1-2
indicates that the biggest differences between the initial
models are in terms of the radial velocity (see also \S 2).
For both metallicity series the ``rotST" models seem to 
collapse with faster speeds. To eliminate this difference,
we re-ran the 1.5-D simulations for the ``norot" and
``rotST" models with zero initial radial velocity and
let {\it FLASH} calculate the collapse velocities
resulting from the dynamical instability in the core.
These simulations yielded somewhat smaller
$^{56}$Ni masses for all models (0.5~$M_{\odot}$ for
200sm\_norot, 1.2~$M_{\odot}$ for 145sm\_rotST, 8.8~$M_{\odot}$
for 245sm\_norot and 9.7~$M_{\odot}$ for 205sm\_rotST) but the
non-monotonicity still remained.

This investigation left us with the option that the source
of non-monotonicity is small initial structural differences in the {\it MESA}
input models that grow in time with the dynamical collapse. Therefore, we 
chose to study the first phases of the 
dynamical collapse of the 1.5-D models in detail
and determine in what portions of the cores significant differences develop
that lead to different final compressions.

Figures 22 and 23 show the results of this dynamical analysis. The initial
acceleration profiles derived by using the {\it MESA} output
are plotted in the top left panels. 
It can already be seen that the inward acceleration in the outer parts of
the CO core ($r \sim$~$2-5 \times 10^{10}$~cm) is somewhat higher
for the ``rotST" models, especially for the $10^{-3}$~$Z_{\odot}$ case.
The top right panels show the radial velocity structure for the initial
phases of the dynamical collapse (0-100~s in steps of 10~s). It is evident
that already at 30-40~s the regions of the core outwards of $\sim$~$2 \times 10^{10}$~cm
collapse somewhat faster for the ``rotST" model, especially for the 10$^{-3}$~$Z_{\odot}$ model series.
This in turn leads to more rapid collapse speeds in the inner regions by $\sim$~80-90~s and, eventually,
the production of a stronger SN shock wave at $\sim$~$10^{10}$~cm.
This suggests that something is different in the dynamical and structural properties
between the ``norot" and the ``rotST" models in the outer core regions ($1-6 \times 10^{10}$~cm).

To further investigate the structural differences between the ``norot" and ``rotST" models
in the outer core regions, we plot their pressure ($P$) and mean molecular weight ($\mu$)
profiles in the lower panels of Figures 22-23. It can be clearly seen that
the ``rotST" models exhibit steeper pressure gradients at $r>1.5 \times 10^{10}$~cm
for the 10$^{-3}$~$Z_{\odot}$ and $r>5 \times 10^{10}$~cm for the
10$^{-4}$~$Z_{\odot}$ model series. On the contrary, the $\mu$-gradients in the same
outer core regions are flatter for the ``rotST" models, an effect that is more clearly visible in the 
10$^{-3}$~$Z_{\odot}$ case and less so in the 10$^{-4}$~$Z_{\odot}$ case. 
We argue that, for the ``rotST" models, the steeper $P$-gradients directly result from
the flatter $\mu$-gradients. To support this argument we consider only the contributions
of gas ($P_{g}=(N/\mu)kT$ where $N$ is the number density and $k$ is the Boltzmann constant) 
and radiation ($P_{r}=a_{r} T^{4}$ where $a_{r}$ is the radiation density constant)
to the total pressure and neglect the effects of e$^{+}$e$^{-}$ pairs
that are less important in the outer core regions where $\Gamma_{1}>$~4/3. We then 
take the gradient of the total pressure:

\begin{equation}
\frac{dP}{dr} = \frac{1}{\mu} kT \frac{dN}{dr} - N k T \frac{1}{\mu^{2}} \frac{d\mu}{dr} + \frac{N}{\mu} k \frac{dT}{dr}
+ 4 a_{r} T^{3} \frac{dT}{dr}.
\end{equation}
This shows that, all else being equal, the steeper the $\mu$-gradient the flatter the $P$-gradient;
exactly the effect we see in Figures 22-23. Consequently, steeper negative $P$-gradients correspond
to higher inward accelerations and radial velocities and more compression in the center.

This result indicates that the source of the non-monotonicity of $^{56}$Ni production is the pre-PISN
rotational mixing that occurs in the ``rotST" models leading to smoother $\mu$-gradients at the CO core - H/He
envelope boundaries. In the ``norot" models, the boundaries are more well defined in terms of $\mu$, and mixing
is minimal yielding a classic onion structure. The steep $\mu$-gradients at core-envelope interfaces lead
to flatter $P$-gradients there and less inward acceleration resulting from the dynamical instability. That
leads to less compression in the core and therefore lower $T_{c}$ values reached, slower nuclear
reaction rates and less $^{56}$Ni mass formed, for the same initial $M_{\rm CO}$.

To further support this result, we ran the ``rotST\_ml2" models for which we adopted a higher
mass loss rate parameter than the standard choice used in the ``rotST" models. High mass-loss
rate can lead to less effective rotational mixing, if the time-scale for mixing is longer compared
to the mass-loss time scale in the outer regions of the progenitor star. This is illustrated by
the $\mu$ profiles derived for the ``rotST\_ml2" models shown as green solid curves in
the lower right panels of Figures 22-23.  The suppression of rotational mixing
leads to steep $\mu$-gradients, similar to those of the ``norot" models for CO cores of the same mass.
That, in turn, also leads to flatter $P$-gradients than in the case of the ``rotST" models and to
lower inward acceleration. For the rapidly-rotating ``rotnoST" models, the angular momentum
barrier that develops is strong enough to counter the effects of steeper $P$-gradients
and to decelerate dynamical collapse.

The experiment with the ``rotST\_ml2" models further supports the idea that the source of the $^{56}$Ni non-monotonicity is the
competition between the effects of rotational mixing in the pre-PISN evolution and the development
of an angular momentum barrier for rapid rotation. 
The effectiveness of the rotational mixing in the pre-PISN {\it MESA}
evolution is, in turn, dependent on other stellar parameters such as metallicity and mass-loss. 
In the 10$^{-4}$~$Z_{\odot}$ model series mass-loss is not as strong as in the 10$^{-3}$~$Z_{\odot}$ case and
rotational mixing is less effective, but the non-monotonicity still pertains. 

\section{SUMMARY AND CONCLUSIONS}

In this work we studied the effects of rotation on PISNe. We used {\it MESA} to calculate
the evolution of massive (135-245~$M_{\odot}$) low metallicity (10$^{-3}$-10$^{-4}$~$Z_{\odot}$)
stars for two different degrees of ZAMS rotation: $\Omega/\Omega_{c} =$~0 and 0.5.
For the rotating models we ran models that include the effects of magnetic fields in the
angular momentum transport and mixing of chemical elements (ST prescriptions) leading
to moderately rotating pre-PISN progenitors (``rotST" models) 
and cases that neglect these effects leading to rapidly rotating progenitors (``rotnoST" models).
We also ran ``rotST" models for which a higher mass loss rate parameter was adopted 
(``rotST\_ml2" models) that led to the suppression of rotational mixing.
The {\it MESA} evolution provided us with PISN progenitor models of the same $M_{\rm CO}$,
for stars in the same metallicity category, that were then mapped conservatively in the AMR
2-D grid of {\it FLASH} including rotation in the direction perpendicular to the grid
(``2.5-D" treatment).
Each PISN explosion was followed with {\it FLASH} until after the SN blast wave
broke out of the stellar surface. Effects such as the mixing, the energetics, and
nucleosynthesis, with emphasis to the production of $^{56}$Ni that powers
the LC of PISNe were studied.

Our 2.5-D simulations revealed mild mixing mainly due to the development of RT instability
in three different regions: at the burning front between the newly formed $^{56}$Ni
and the $^{16}$O shell, at the interface between the $^{16}$O shell
and the He layer and, lastly, due to the reverse shock that develops
after the SN shock wave breaks out of the stellar surface into a circumstellar wind material. 
Rapid rotation also yielded modest explosion asymmetries due to the development of
a strong angular momentum barrier: regions close to the equator collapsed
with lower speeds than polar ones. This behavior is in agreement with the findings
of Glatzel et al. (1985) who also determined larger acceleration along the symmetry axis using
their Maclaurin spheroids method.
Both the SN blast wave and the
distribution of $^{56}$Ni were asymmetric for the rapidly rotating ``rotnoST" models.

We found a non-monotonic production of $^{56}$Ni with increased
rotation: non-rotating models produced less $^{56}$Ni than slowly-rotating ``rotST" models
while rapidly rotating ``rotnoST" CO cores produced the smallest amounts amongst
all models, all for the same initial $M_{\rm CO}$. 
We determined that the source of the $^{56}$Ni production non-monotonicity is
the competition between the effects of pre-PISN rotational mixing in the 
stratification of chemical species in the core and the development of
strong centrifugal forces that counter collapse. 
Effective rotational mixing leads to smooth $\mu$-gradients in the
CO core - H/He envelope interface that, in turn, form steep
pressure gradients corresponding to higher inward acceleration
following the e$^{+}$e$^{-}$ pair dynamical instability. This
is in contrast with the
case of no rotation where the progenitor star forms a classic
onion structure with well-defined boundaries between layers of
different composition. Higher inward acceleration leads 
to more compression in the core, higher central temperatures,
increased nuclear reaction rates and, eventually, larger
amounts of $^{56}$Ni produced that can power the SN
LC. 

The non-monotonic behavior vanishes when the
effects of rotational mixing are suppressed (for example, due to
increased mass-loss rates preventing effective mixing in the outer
parts of the CO cores). For PISN progenitors of
the same $M_{\rm CO}$ and identical structural characteristics
increased rotation leads to less energetic explosions that
produce less $^{56}$Ni. This result is in good agreement
with the findings of Glatzel et al. (1985).

Our work is the first in the literature to present 2-D simulations of PISNe with
rotation. Our results indicate that the final LCs and spectral
characteristics of these events are expected to have only modest dependence
on initial rotation rates because the differences
in the final $^{56}$Ni produced are not significant, at least
in the case of moderate rotation. 
In the case of high rotation rates of the CO core, the
PISN explosions are expected to be dim and red due to the
small amount of $^{56}$Ni produced, and probably
to have a long duration LC due to photon diffusion through
a large SN ejecta mass.
If the first massive stars formed in the Universe after
the end of the Dark Ages are rapid rotators as predicted in 
several simulations (Greif et al. 2011; Stacy et al. 2013), 
then their PISN explosions that missions
such as the {\it JWST} and {\it WFIRST} aim
to discover may be somewhat redder than in the
case of zero rotation due to the decreased $^{56}$Ni
mass. 
Given that the effectiveness of rotational
mixing in the pre-PISN evolution of low-metallicity
stars is unconstrained and dependent upon
several stellar parameters, it will be hard to predict
the rotational speed of the progenitor star
based on the PISN observations alone.

\acknowledgments

We would like to thank David Dearborn, Christopher Lindner, Daniel Whalen, Matteo Cantiello
and the anonymous referee
for useful discussions and comments. We also thank the Texas Advanced Computing
Center for providing us with a generous allocation of computation time on the {\it Stampede} 
supercomputer. This research is supported by the
STScI grant AR12820. EC would like to thank the University of Texas Graduate
School William C. Powers fellowship for its support of his studies. 
Some work on this paper by JCW was done in the hospitable clime of the Aspen
Center for Physics that is supported by the NSF Grant PHY-1066293.


{}                     

\clearpage
\begin{figure}
\begin{center}
\includegraphics[angle=0,width=16cm]{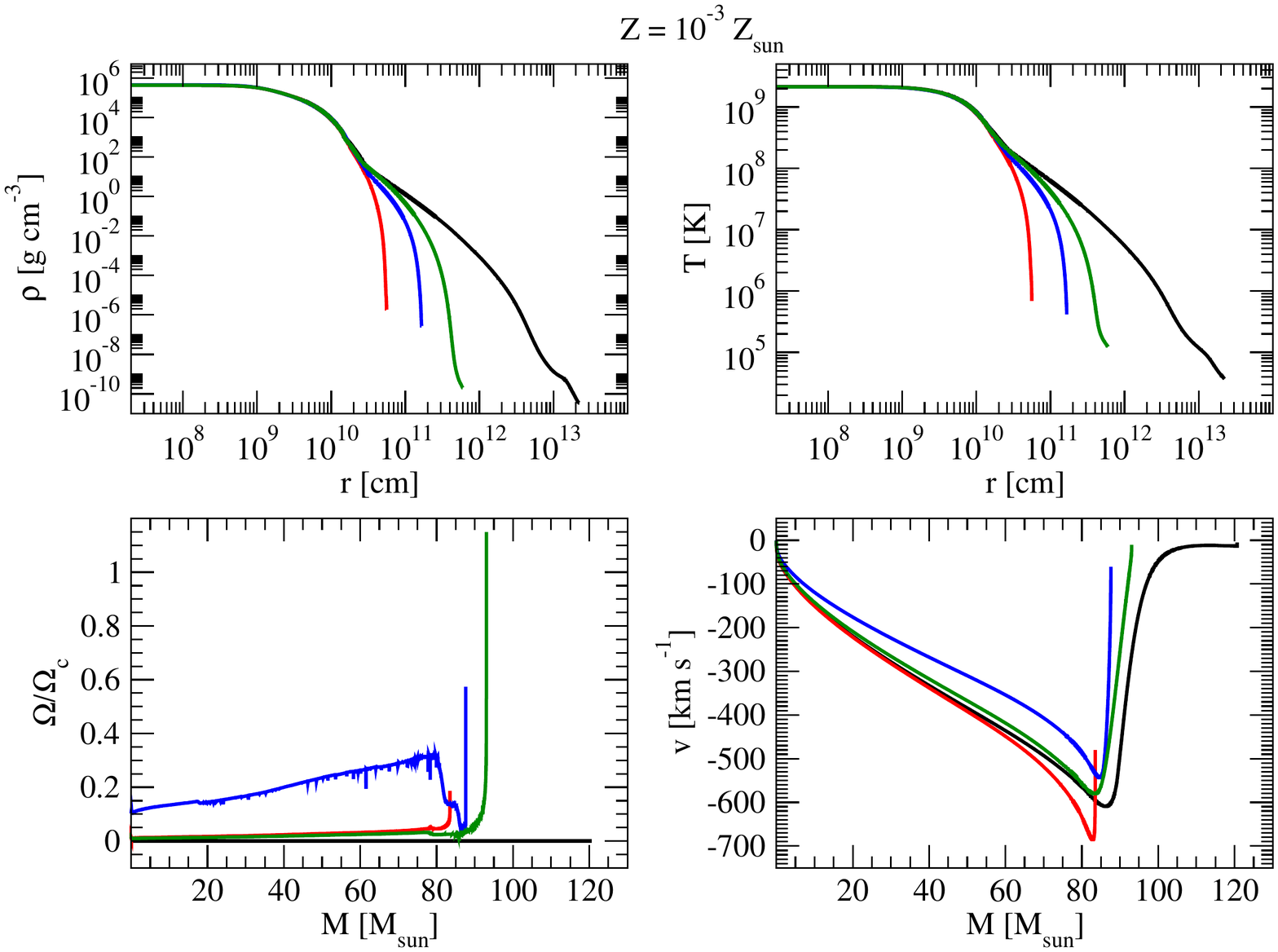}
\caption{Internal structure plots for the $Z =$~$10^{-3}$~$Z_{\odot}$ {\it MESA} PISN input models at the time of mapping to {\it FLASH}. The density and temperature structures are shown in the upper panels while the internal angular velocity profiles in terms of the critical value, are given in the lower left panel and the radial velocity profiles in the lower right panel.
Black curves represent the non-rotating (``norot") models, red curves the rotating models with the ST effects
included (``rotST"), blue curves the rotating models without the effects of ST (``rotnoST") and green curves the rotating models with the
ST effects included but with a higher mass-loss rate used in the evolution (``rotST\_ml2").}
\end{center}
\end{figure}

\begin{figure}
\begin{center}
\includegraphics[angle=0,width=16cm]{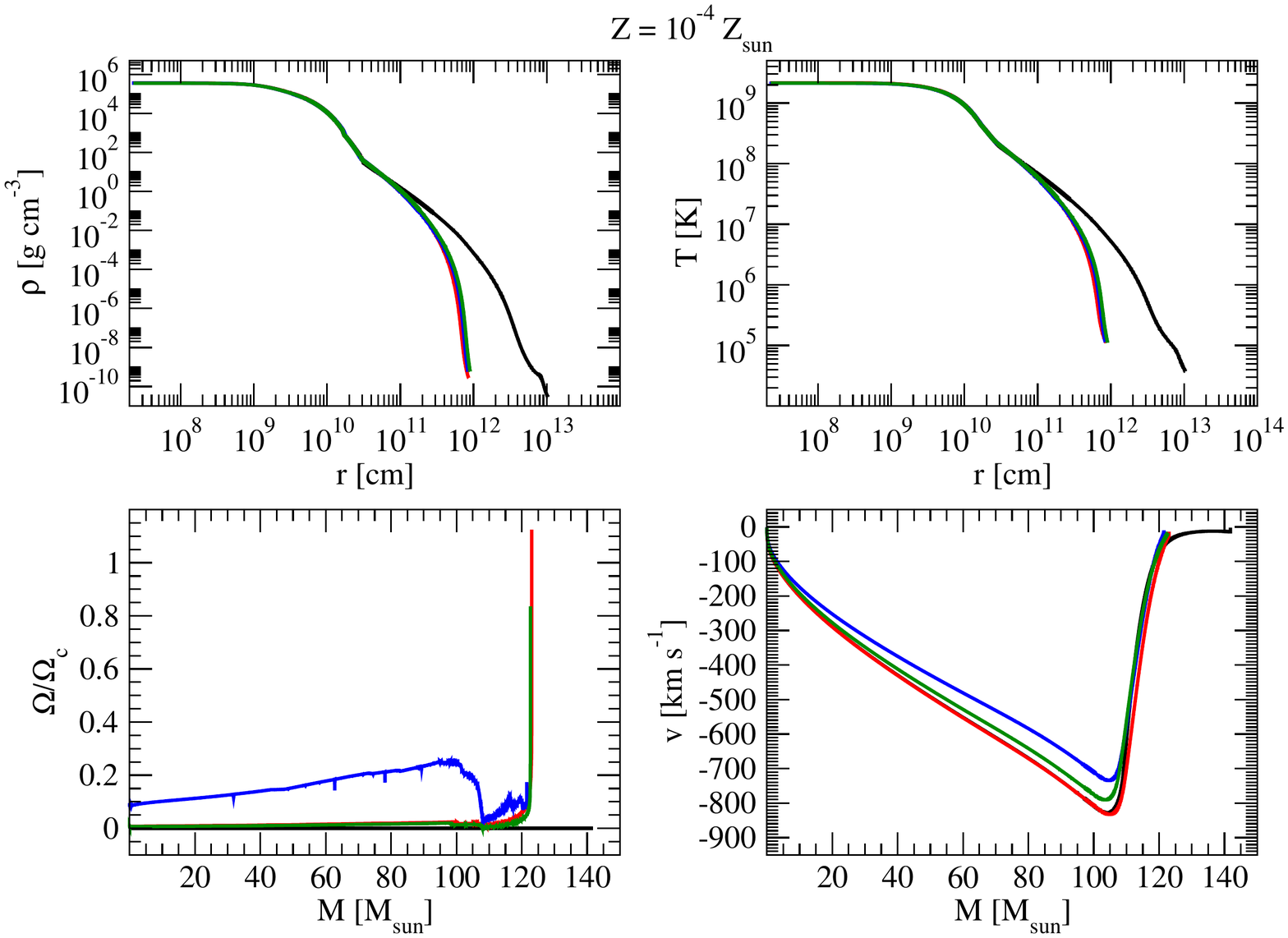}
\caption{Same as Figure 1 but for the $Z =$~$10^{-4}$~$Z_{\odot}$ {\it MESA} PISN input models.}
\end{center}
\end{figure}

\begin{figure}
\begin{center}
\includegraphics[angle=0,width=16cm]{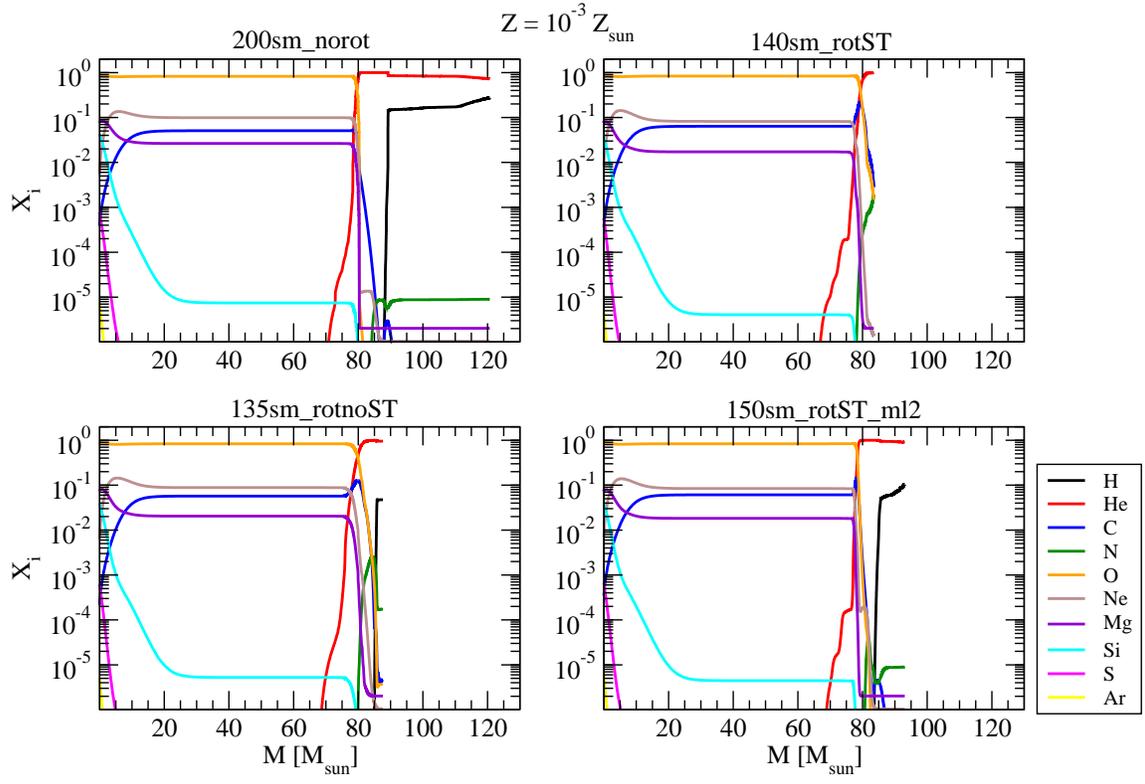}
\caption{Compositional structures for the $Z =$~$10^{-3}$~$Z_{\odot}$  {\it MESA} PISN 
input models at the time of mapping to {\it FLASH}.}
\end{center}
\end{figure}

\begin{figure}
\begin{center}
\includegraphics[angle=0,width=16cm]{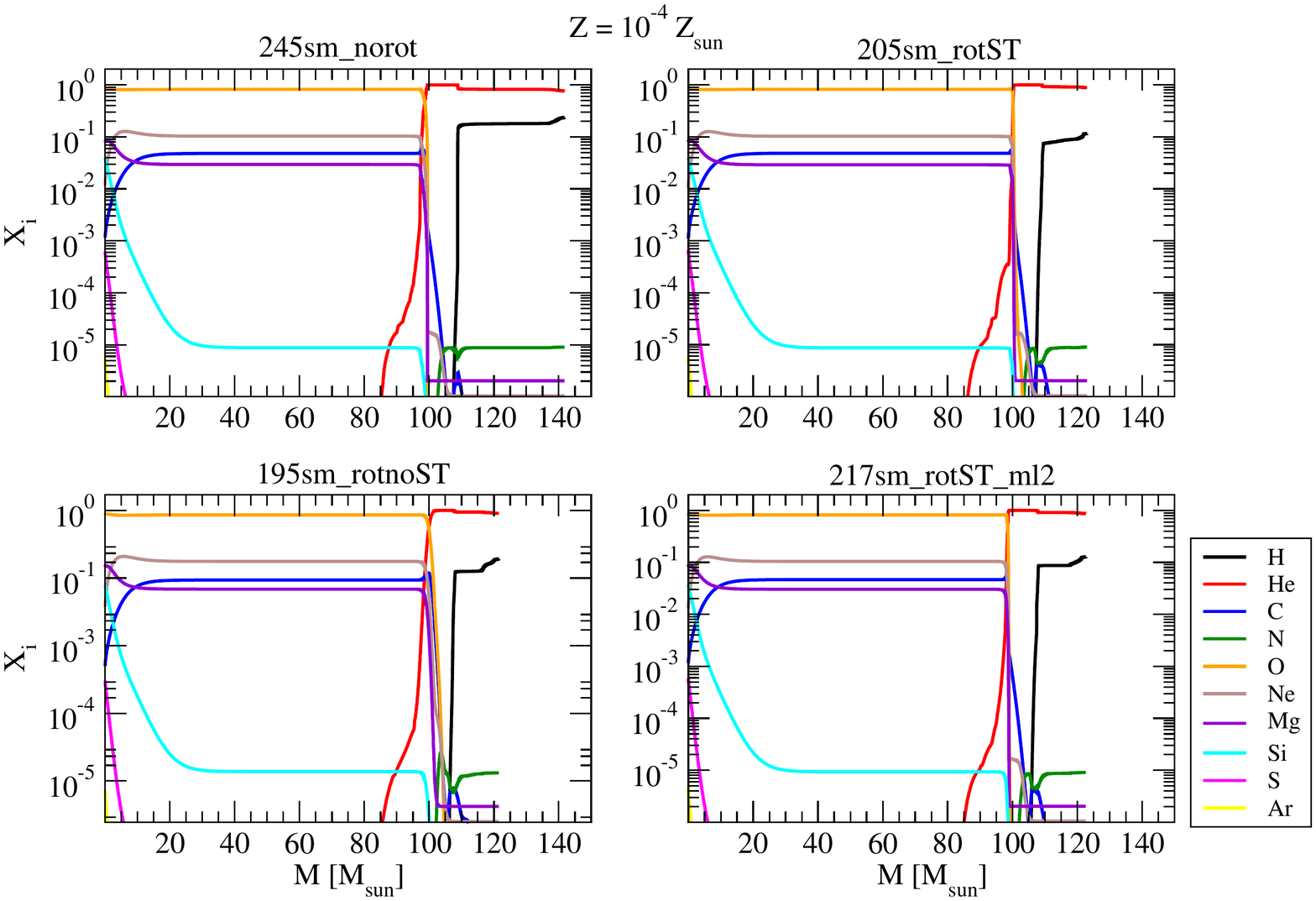}
\caption{Same as Figure 3 but for the $Z =$~$10^{-4}$~$Z_{\odot}$ {\it MESA} PISN input models.}
\end{center}
\end{figure}

\begin{figure}
\begin{center}
\includegraphics[angle=0,width=16cm]{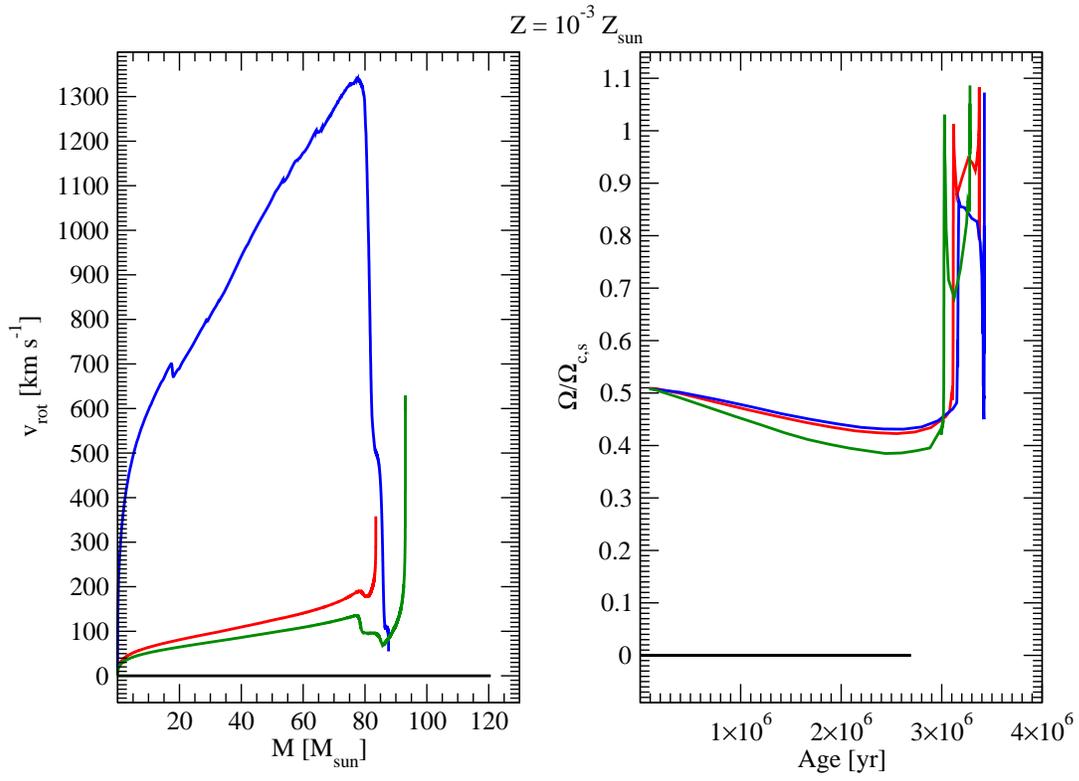}
\caption{Internal rotational profiles (left panel) and evolution of $\Omega/\Omega_{c,s}$
(right panel) for the $Z =$~$10^{-3}$~$Z_{\odot}$ {\it MESA} PISN input models.}
\end{center}
\end{figure}

\begin{figure}
\begin{center}
\includegraphics[angle=0,width=16cm]{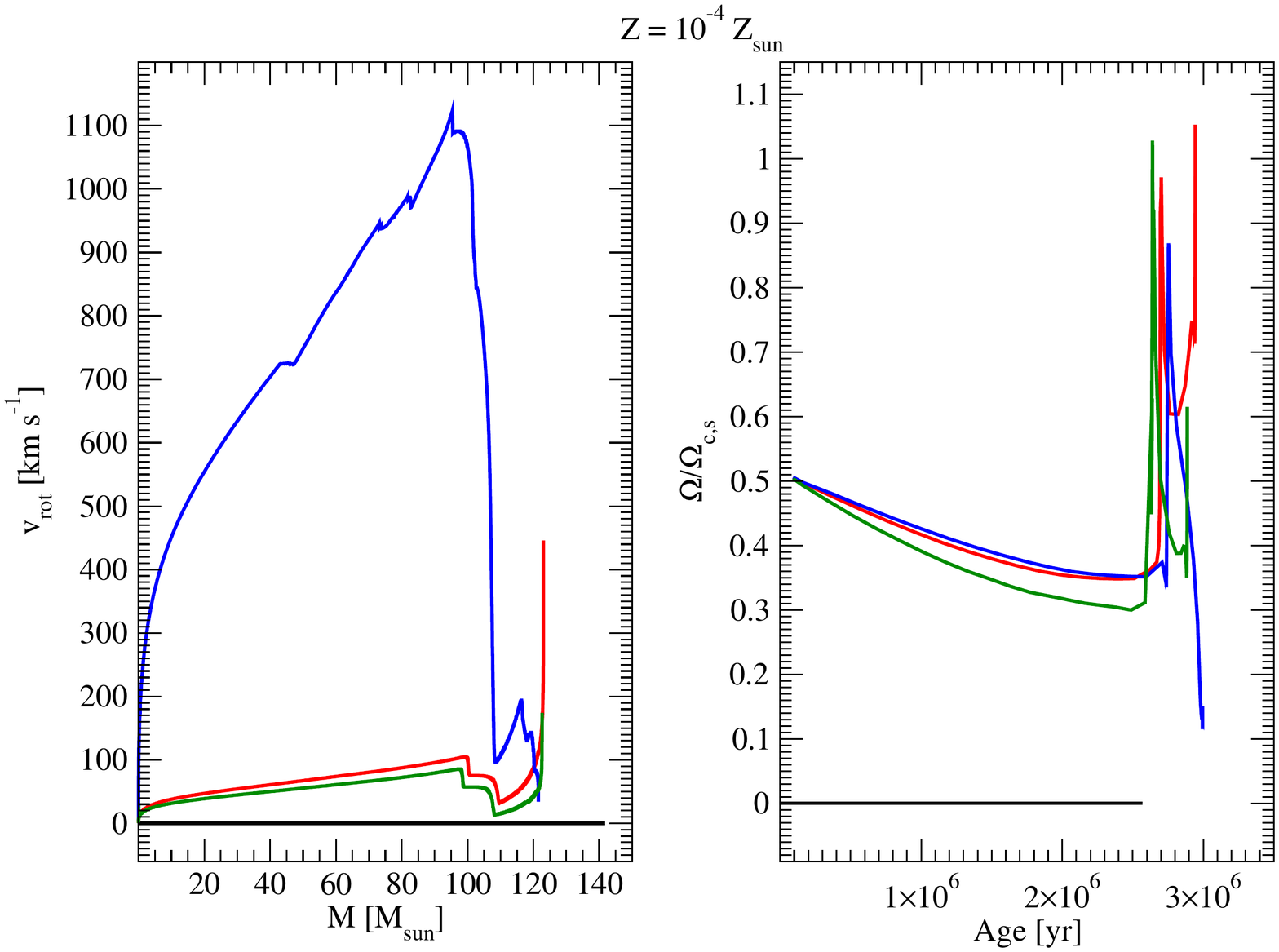}
\caption{Same as Figure 5 but or the $Z =$~$10^{-4}$~$Z_{\odot}$ {\it MESA} PISN input models.}
\end{center}
\end{figure}

\begin{figure}
\begin{center}
\includegraphics[angle=0,width=14cm]{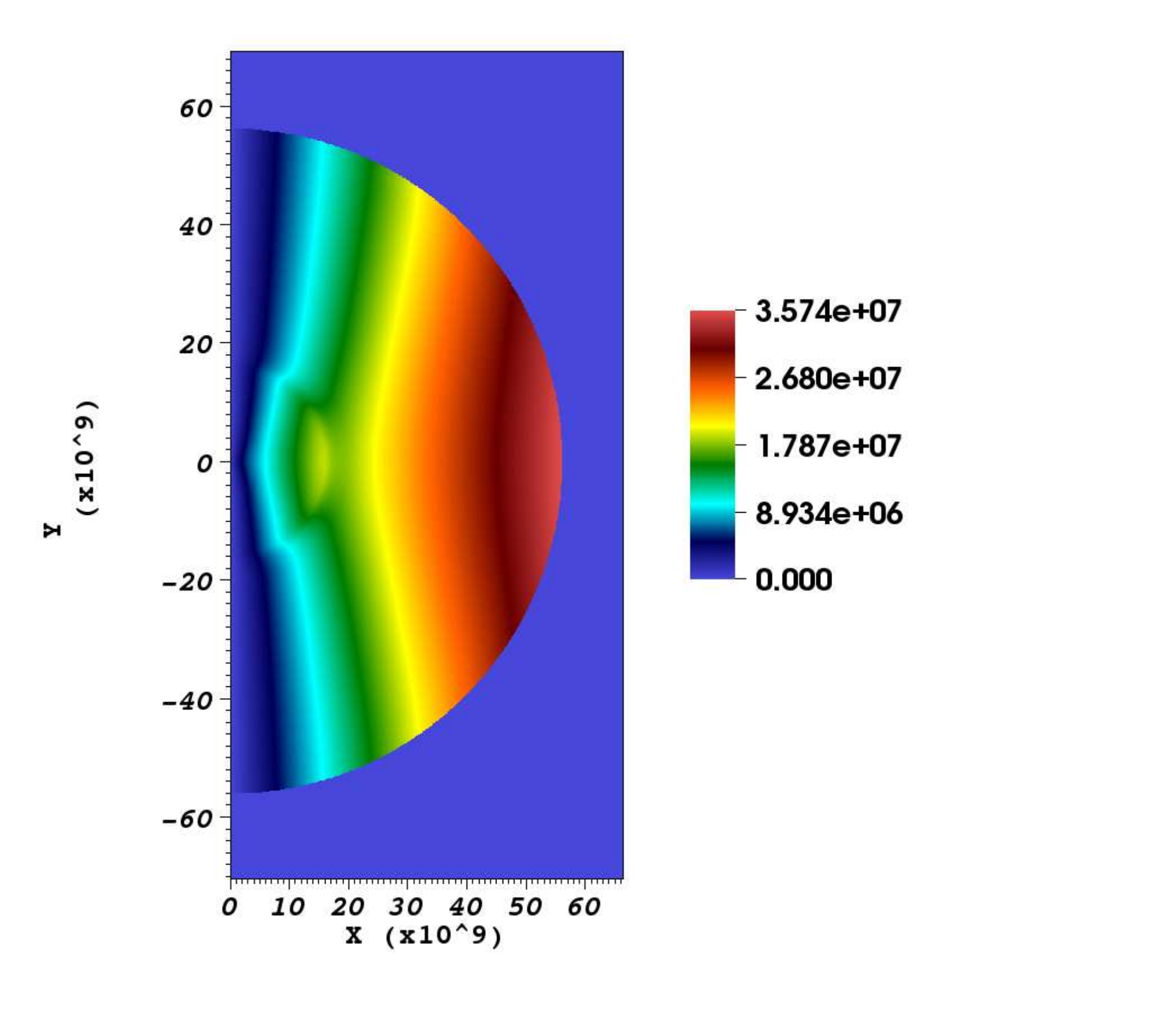}
\caption{The initial rotation velocity profile of the {\it MESA} 
PISN progenitor model 140sm\_rotST mapped in the 2-D grid of 
{\it FLASH} assuming shellular rotation. The polar axis (indicated as ``Y-axis" in the Figure) 
is coincident with the axis of rotation and the rotational 
velocity vectors point into the plane of the plot. The color-coded legend shows the values for the 
rotational velocity in cm~s$^{-1}$.}
\end{center}
\end{figure}

\begin{figure}       
\centerline{
\hskip -0.2 in
\psfig{figure=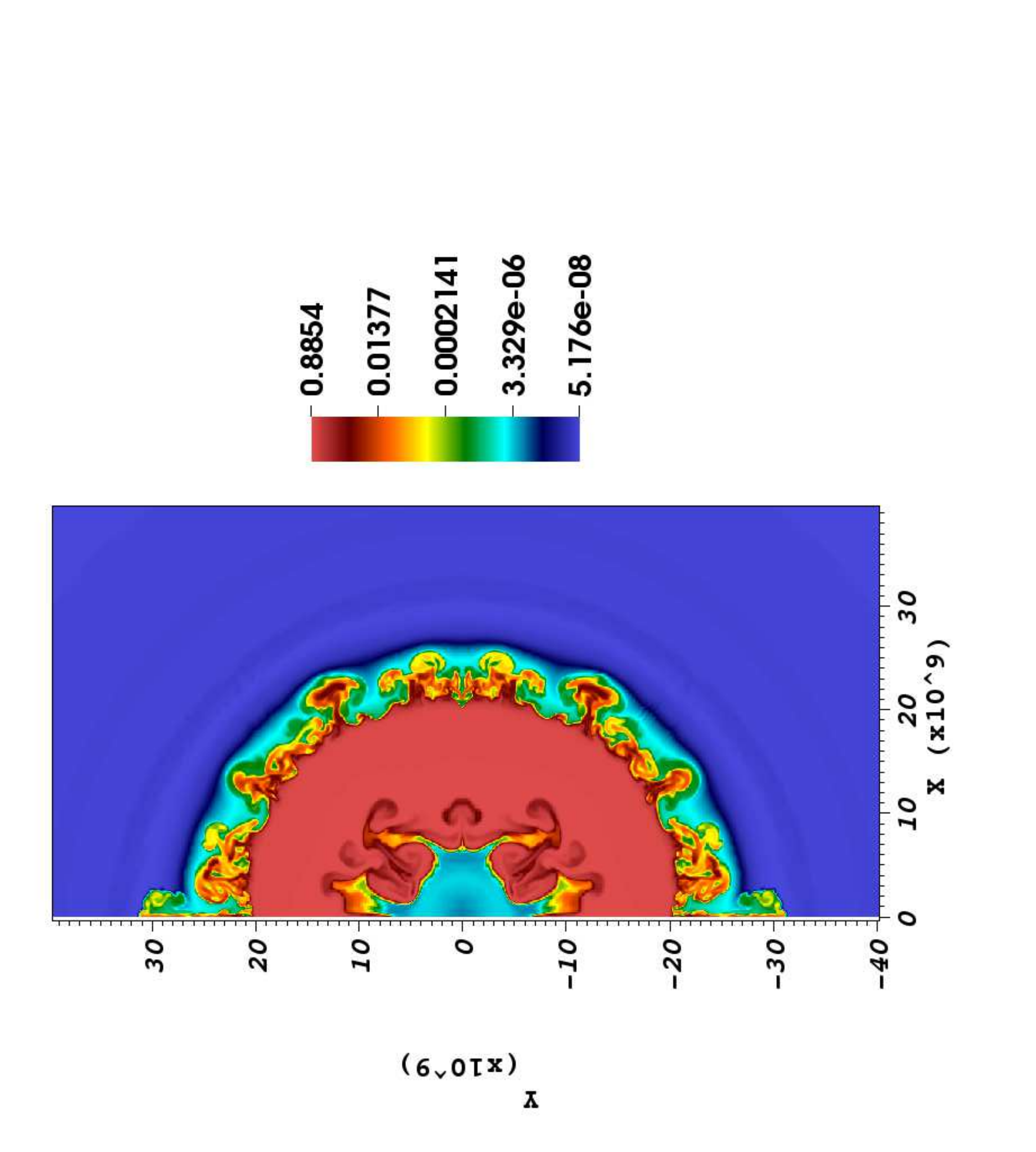,angle=-90,width=4in}
\psfig{figure=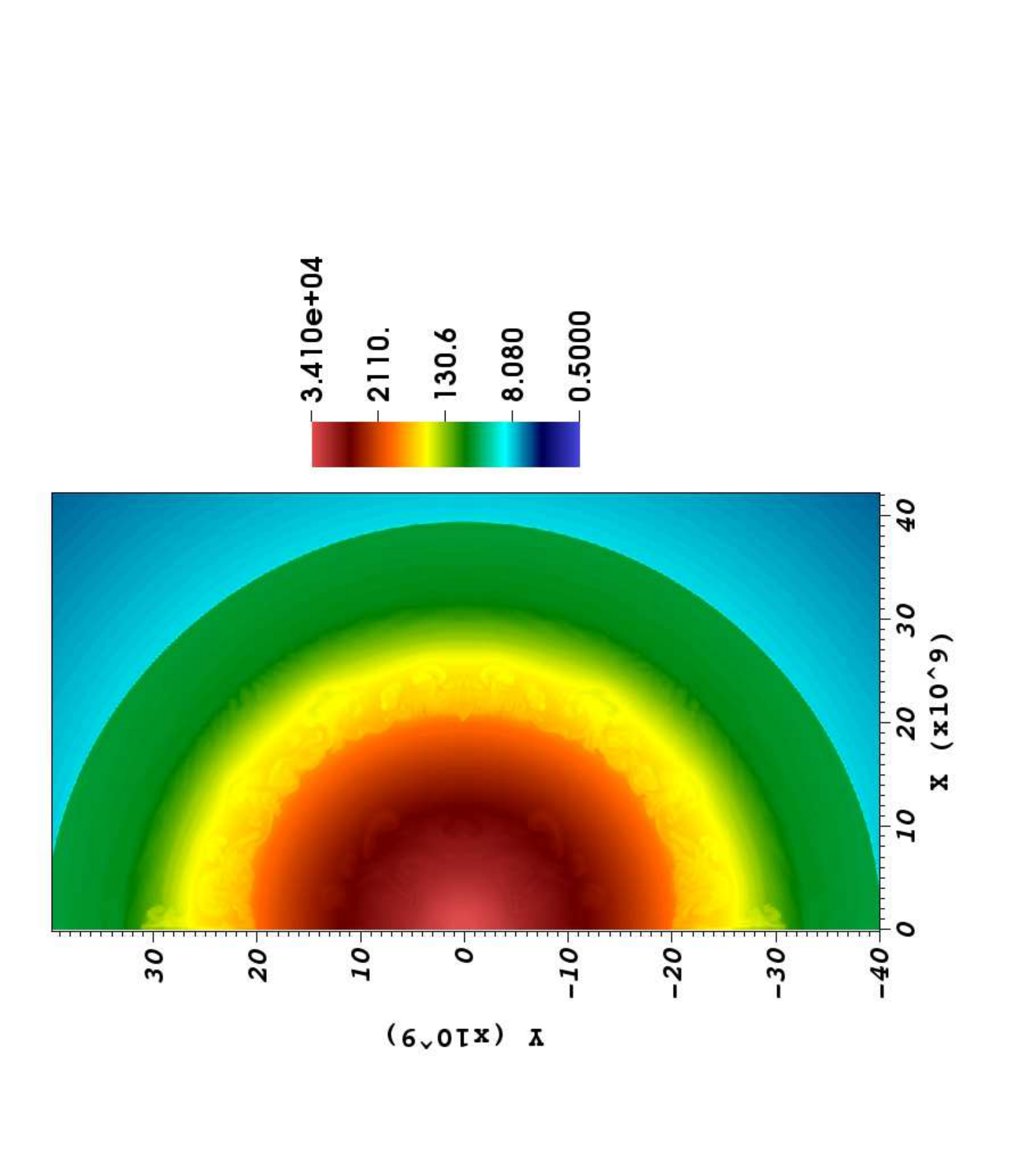,angle=-90,width=4in}
}
\caption{{\it Left Panel}:  
The $^{16}$O mass fraction of the PISN model 200sm\_norot ($Z =$~$10^{-3}$~$Z_{\odot}$) at time $t =$~ 103~s after the onset of
dynamical collapse.
{\it Right Panel}: The density profile of the same model at the same instant. The color-coded legends show $^{16}$O mass 
fraction and density in units of g~cm$^{-3}$.}
\end{figure}

\begin{figure}       
\centerline{
\hskip -0.2 in
\psfig{figure=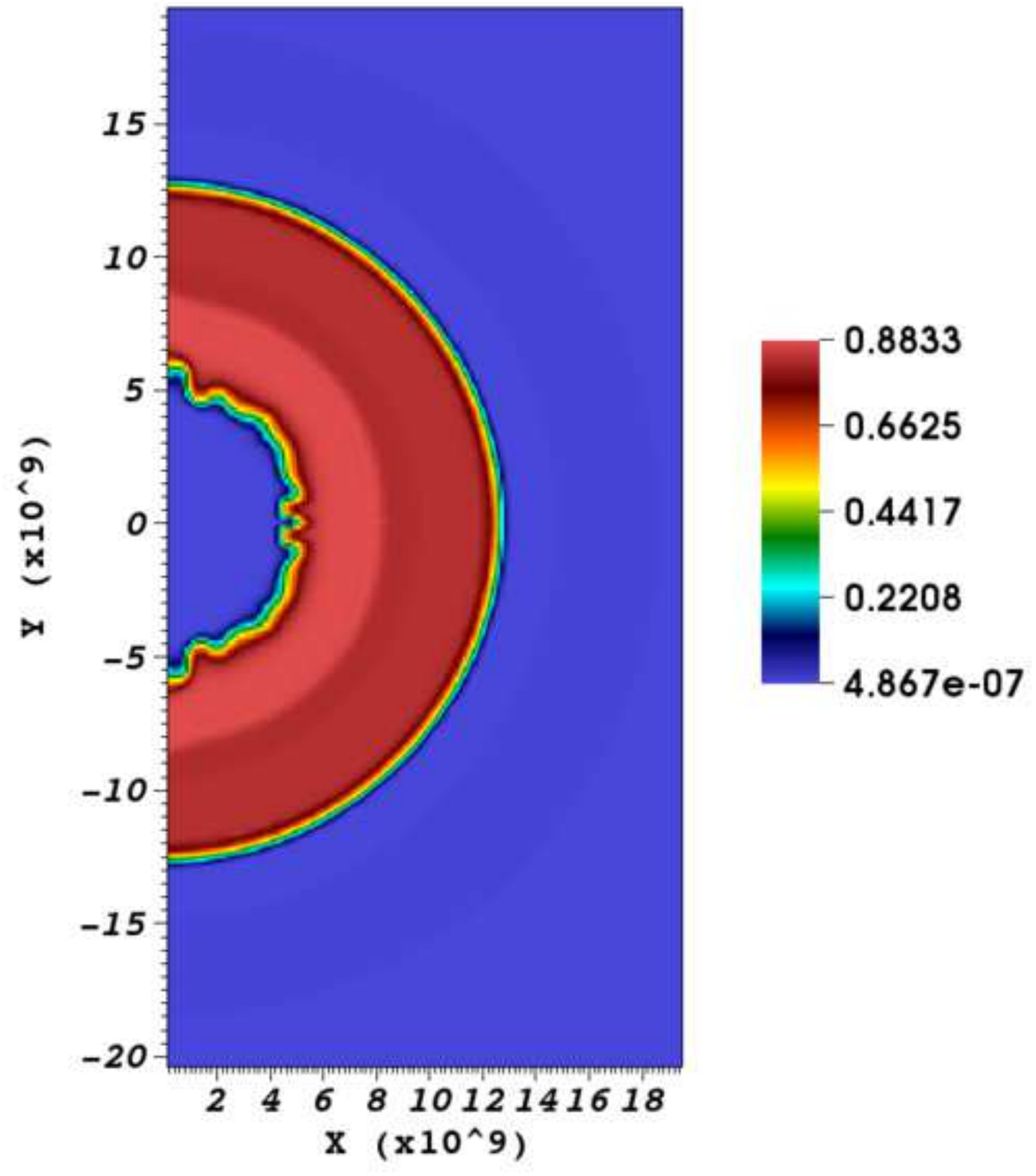,angle=0,width=4in}
\psfig{figure=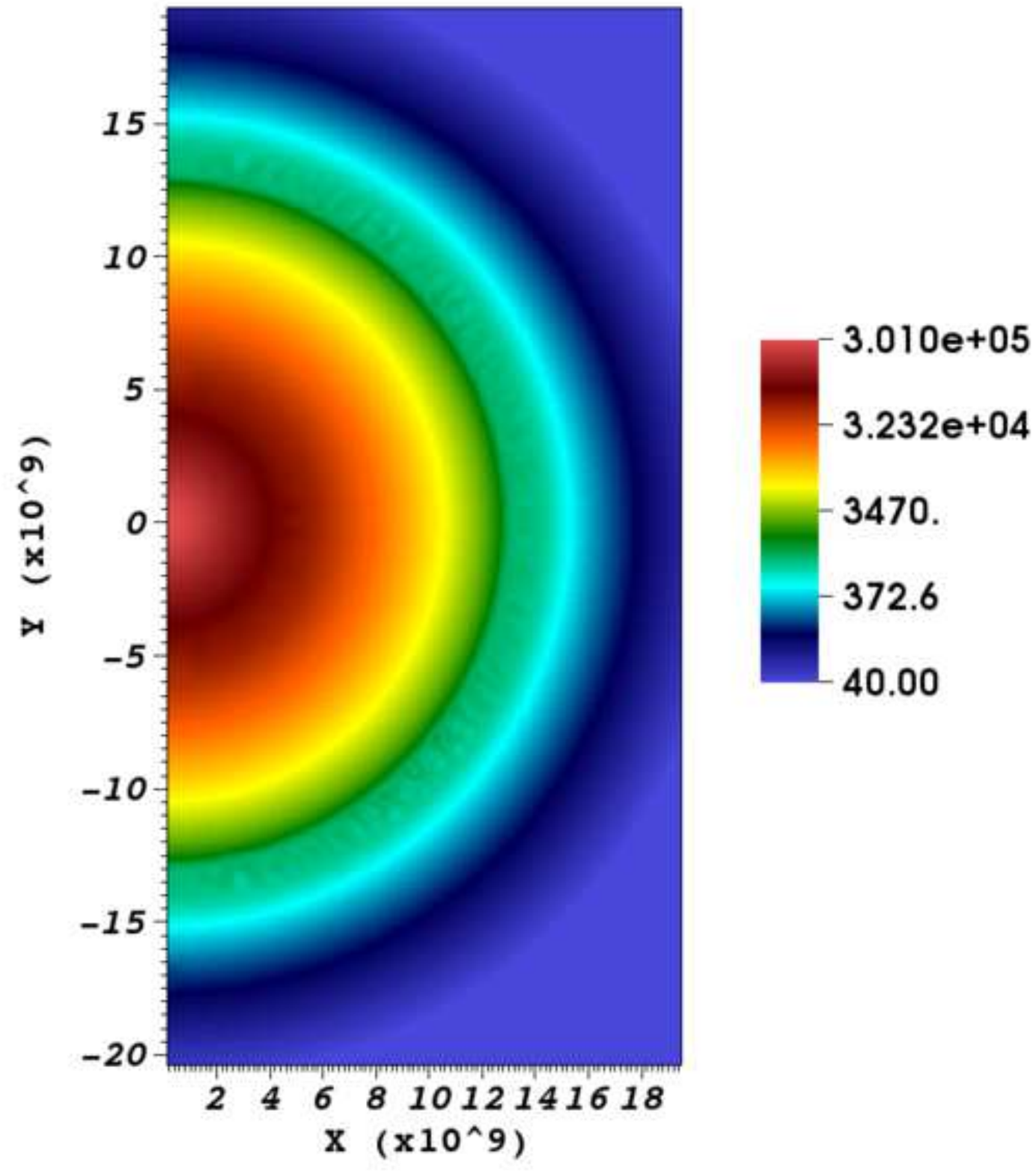,angle=0,width=4in}
}
\caption{Same as Figure 8 but for PISN model 140sm\_rotST ($Z =$~$10^{-3}$~$Z_{\odot}$) at time $t =$~ 82~s after 
the onset of dynamical collapse.}
\end{figure}

\begin{figure}       
\centerline{
\hskip -0.2 in
\psfig{figure=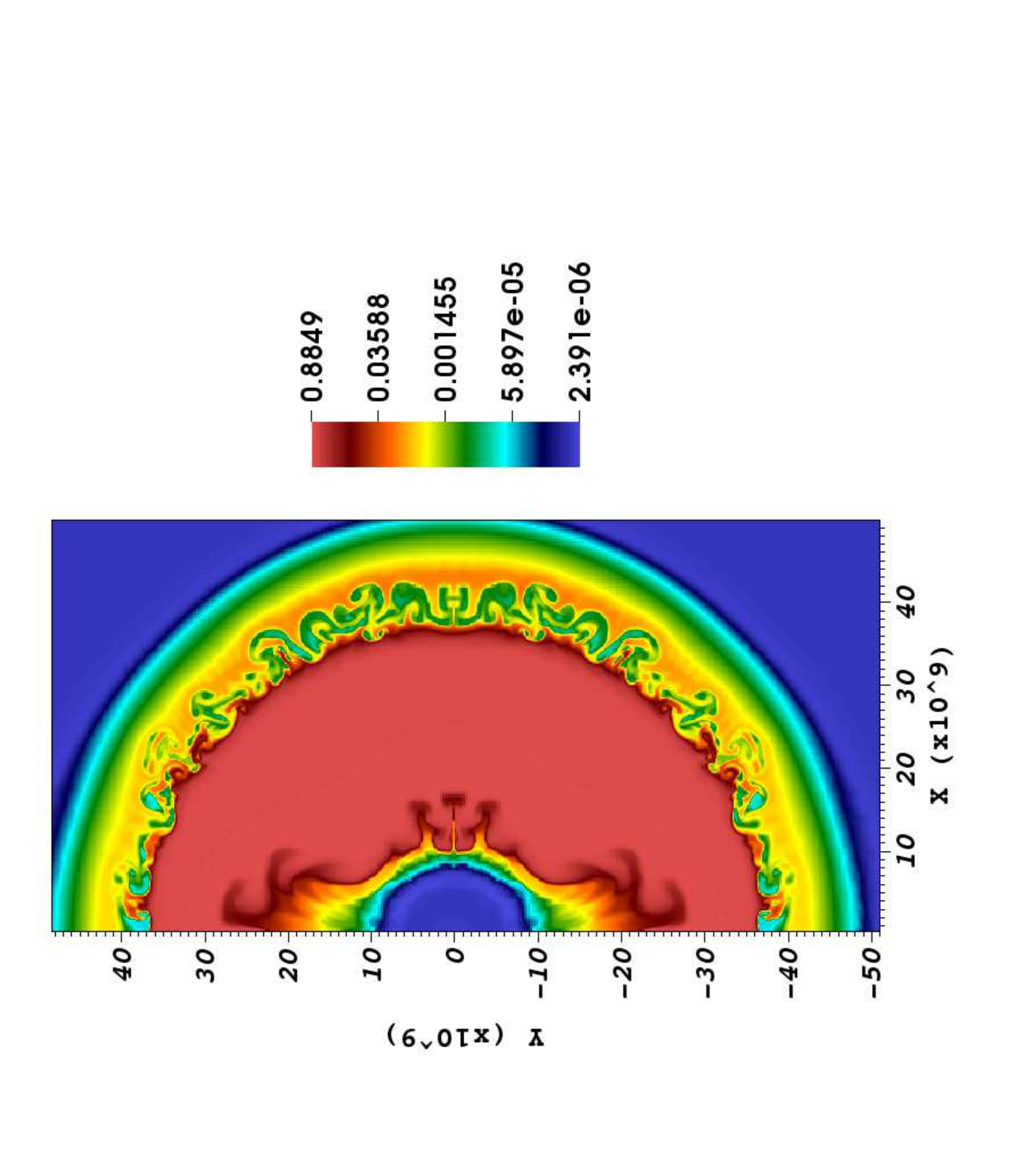,angle=-90,width=4in}
\psfig{figure=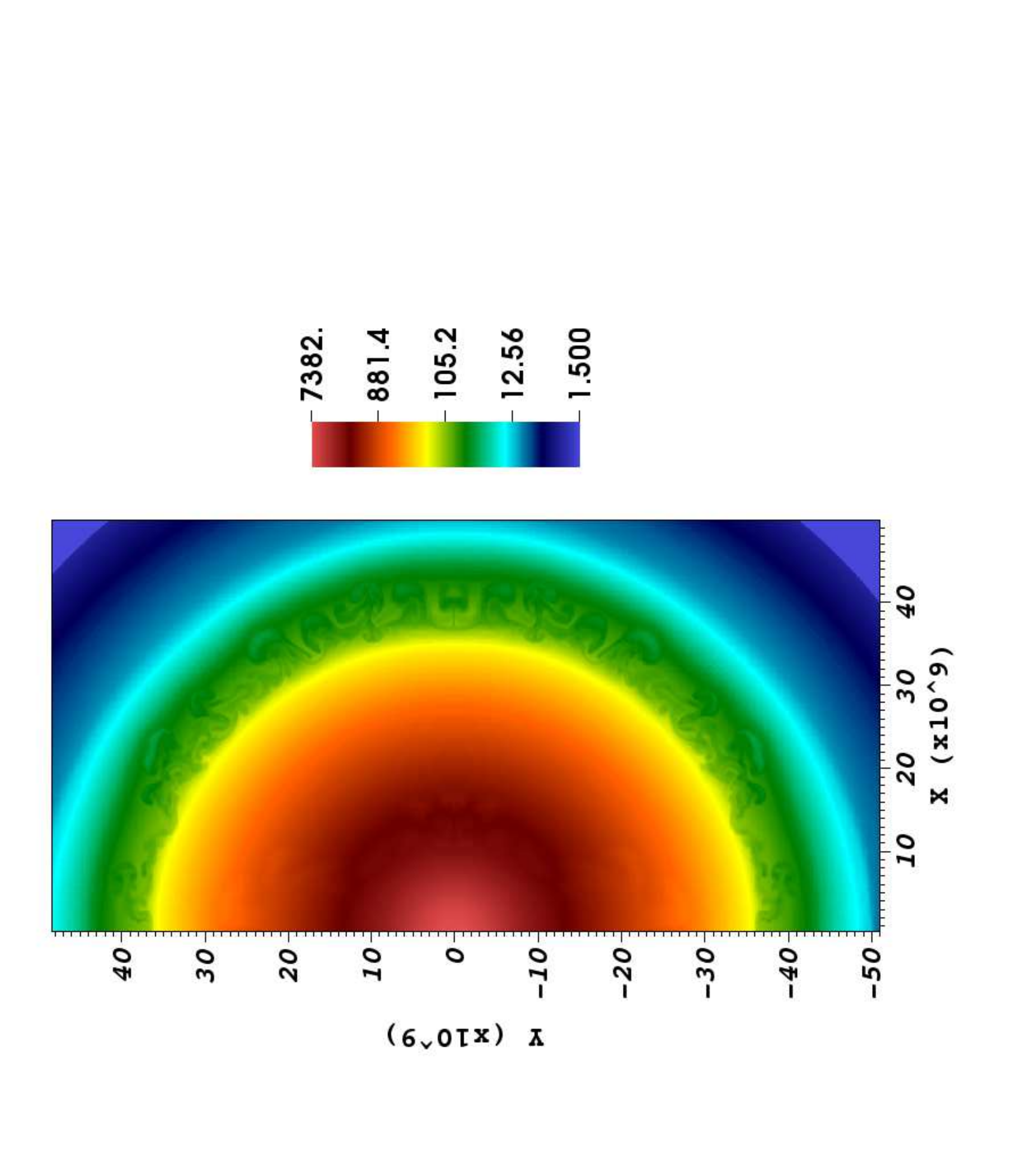,angle=-90,width=4in}
}
\caption{Same as Figure 8 but for PISN model 135sm\_rotnoST ($Z =$~$10^{-3}$~$Z_{\odot}$) at time $t =$~ 138~s after 
the onset of dynamical collapse.}
\end{figure}

\begin{figure}       
\centerline{
\hskip -0.2 in
\psfig{figure=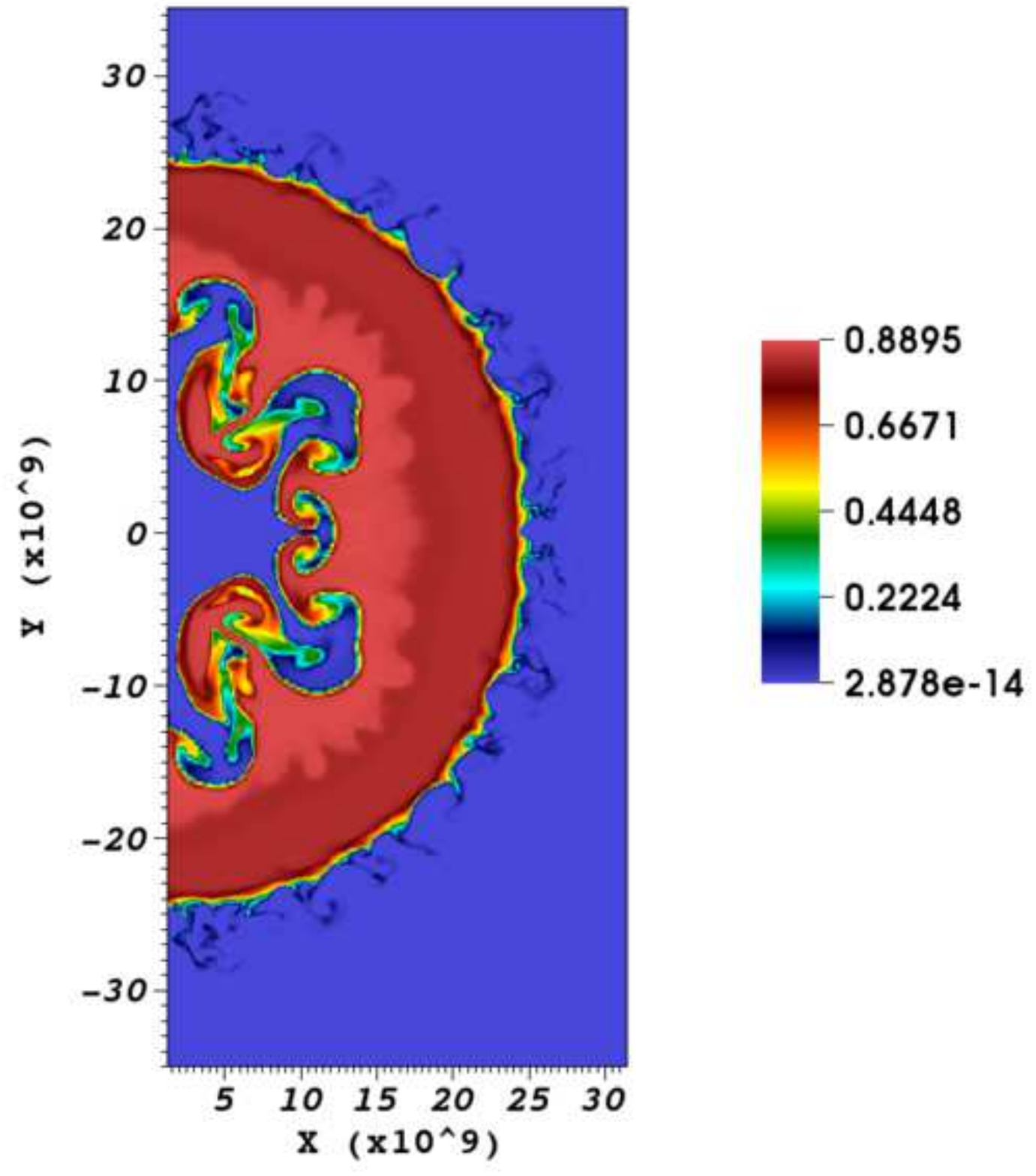,angle=0,width=4in}
\psfig{figure=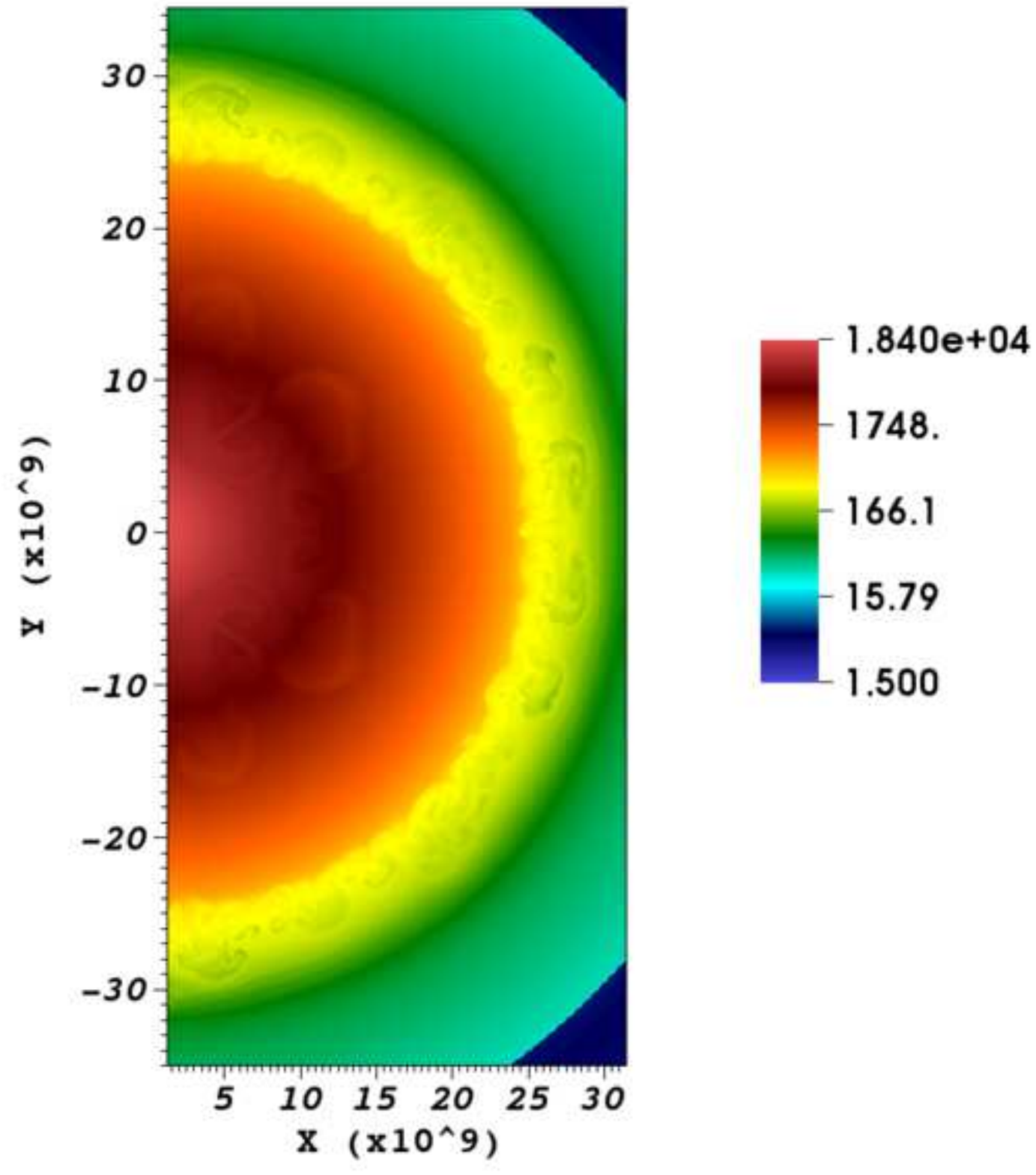,angle=0,width=4in}
}
\caption{Same as Figure 8 but for PISN model 150sm\_rotST\_ml2 ($Z =$~$10^{-3}$~$Z_{\odot}$) at time $t =$~ 112~s after 
the onset of dynamical collapse.}
\end{figure}

\begin{figure}       
\centerline{
\hskip -0.2 in
\psfig{figure=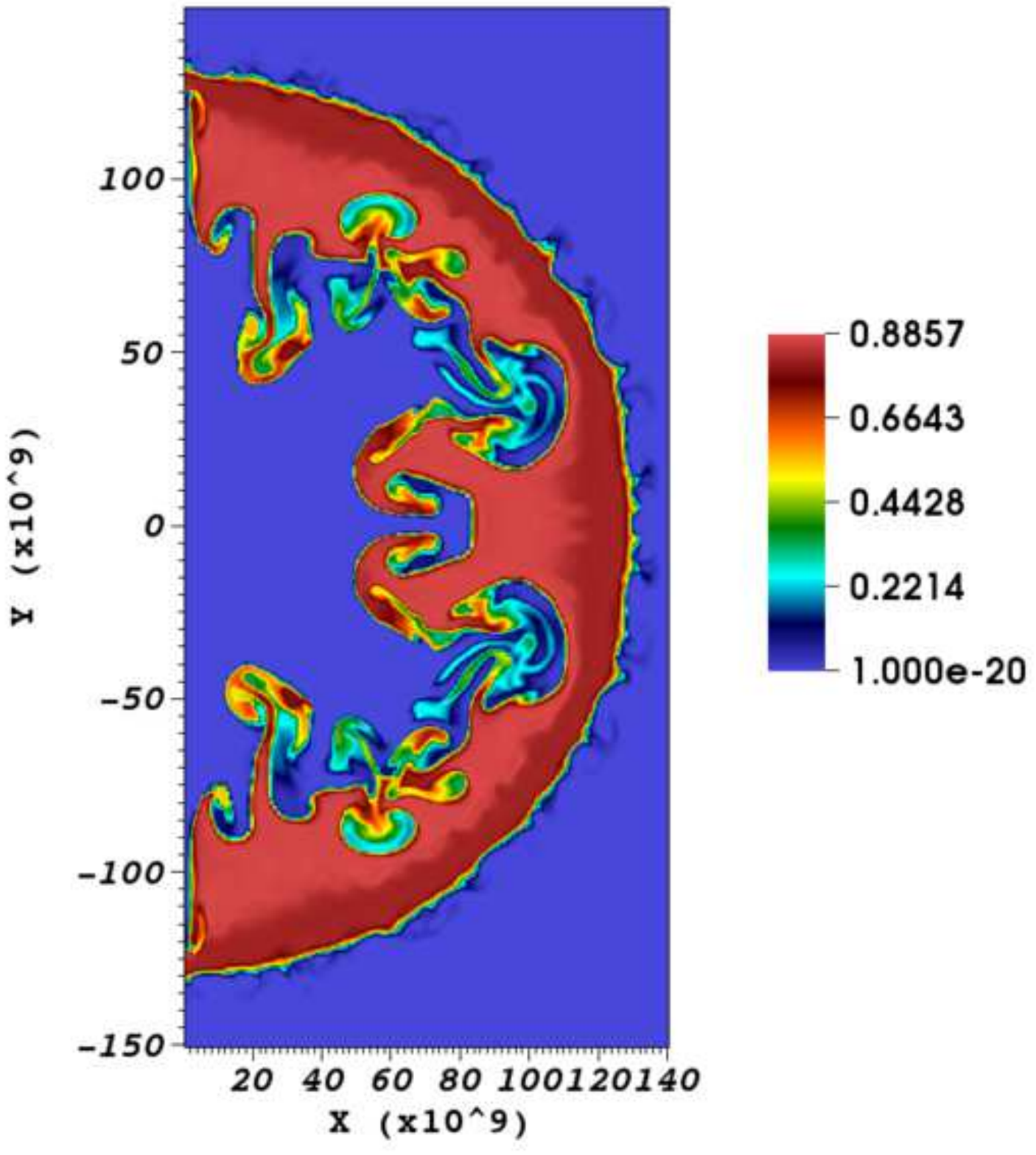,angle=0,width=4in}
\psfig{figure=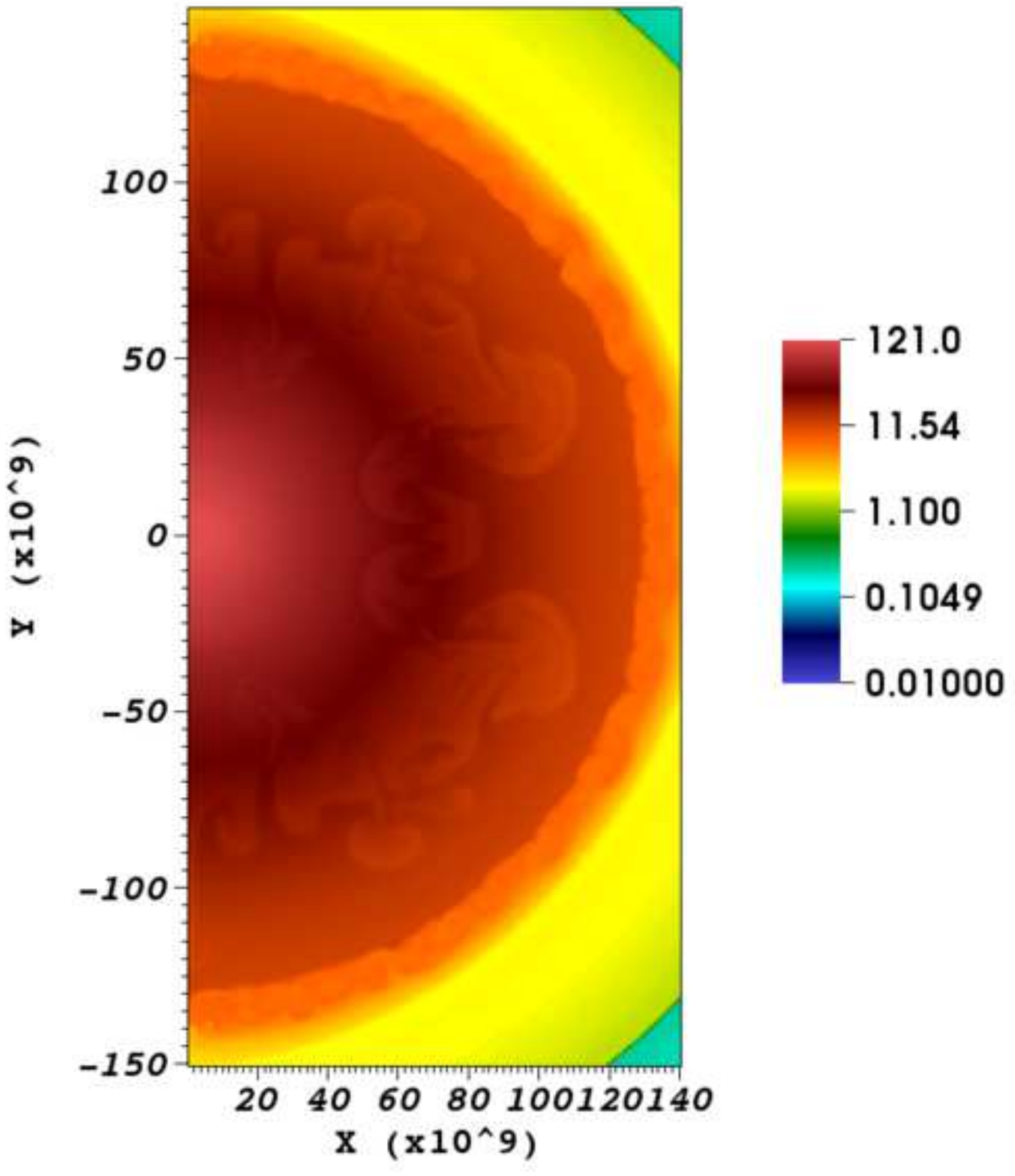,angle=0,width=4in}
}
\caption{Same as Figure 8 but for PISN model 245sm\_norot ($Z =$~$10^{-4}$~$Z_{\odot}$) at time $t =$~ 245~s after 
the onset of dynamical collapse.}
\end{figure}

\begin{figure}       
\centerline{
\hskip -0.2 in
\psfig{figure=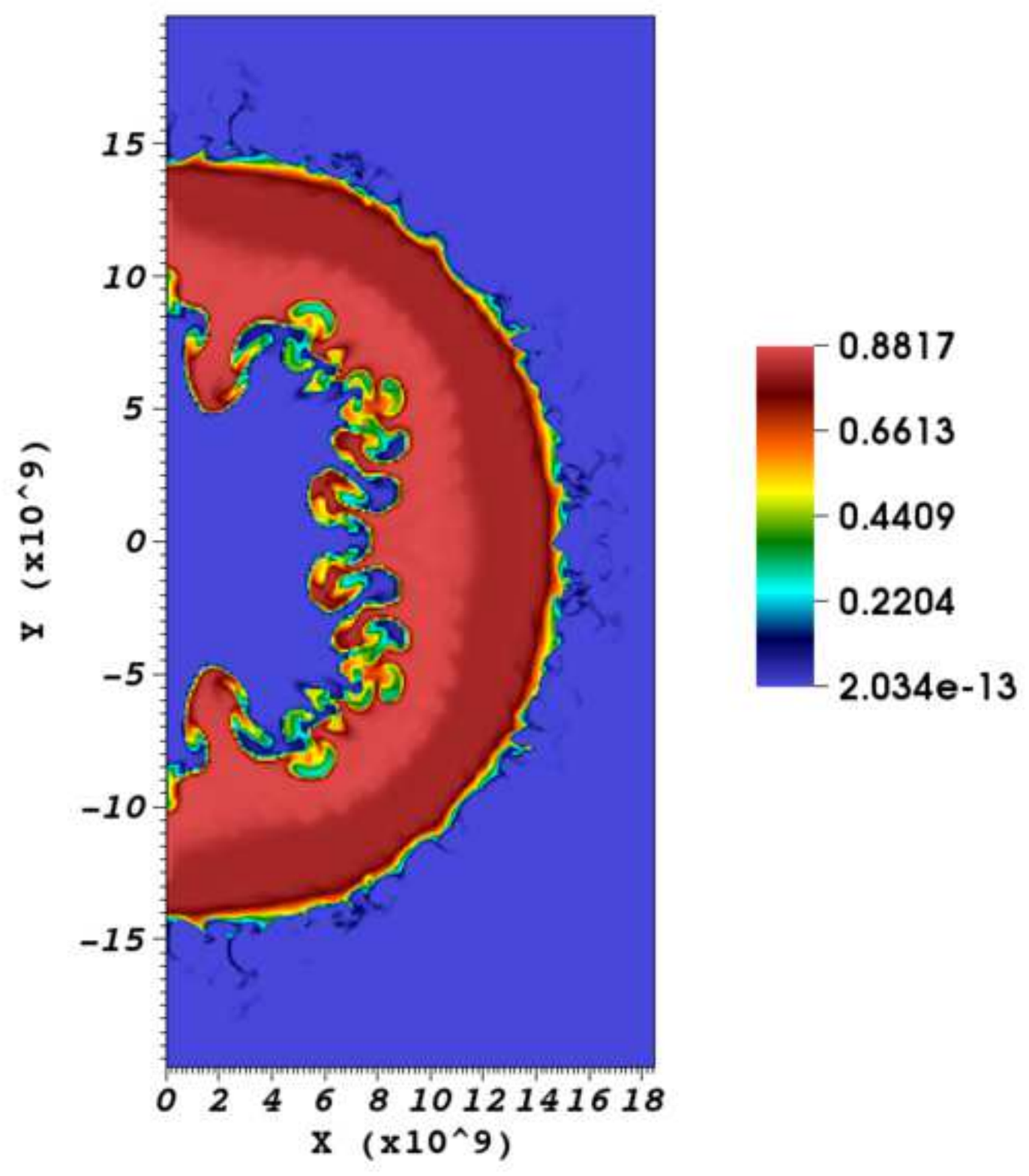,angle=0,width=4in}
\psfig{figure=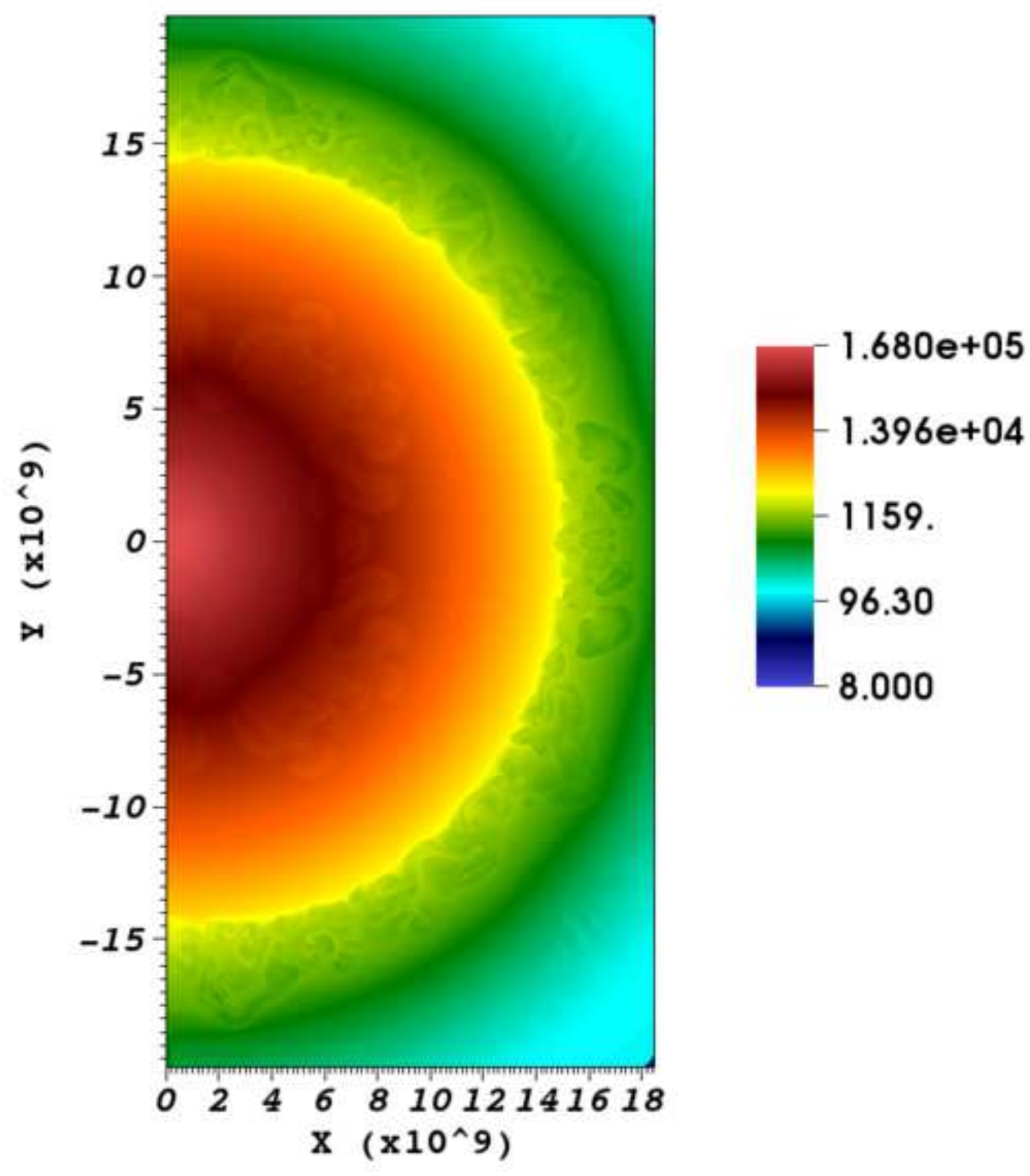,angle=0,width=4in}
}
\caption{Same as Figure 8 but for PISN model 205sm\_rotST ($Z =$~$10^{-4}$~$Z_{\odot}$) at time $t =$~ 81~s after 
the onset of dynamical collapse.}
\end{figure}

\begin{figure}       
\centerline{
\hskip -0.2 in
\psfig{figure=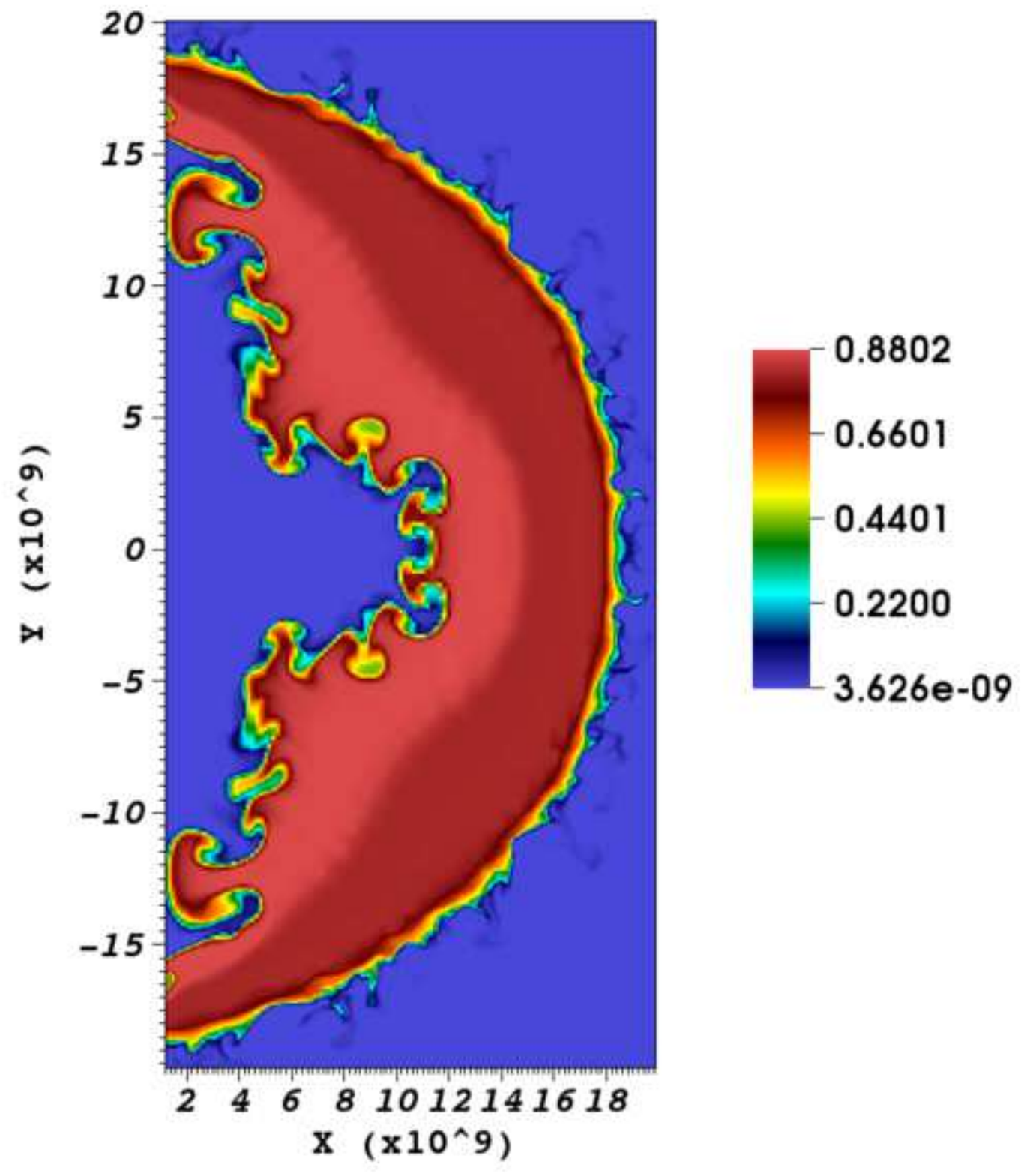,angle=0,width=4in}
\psfig{figure=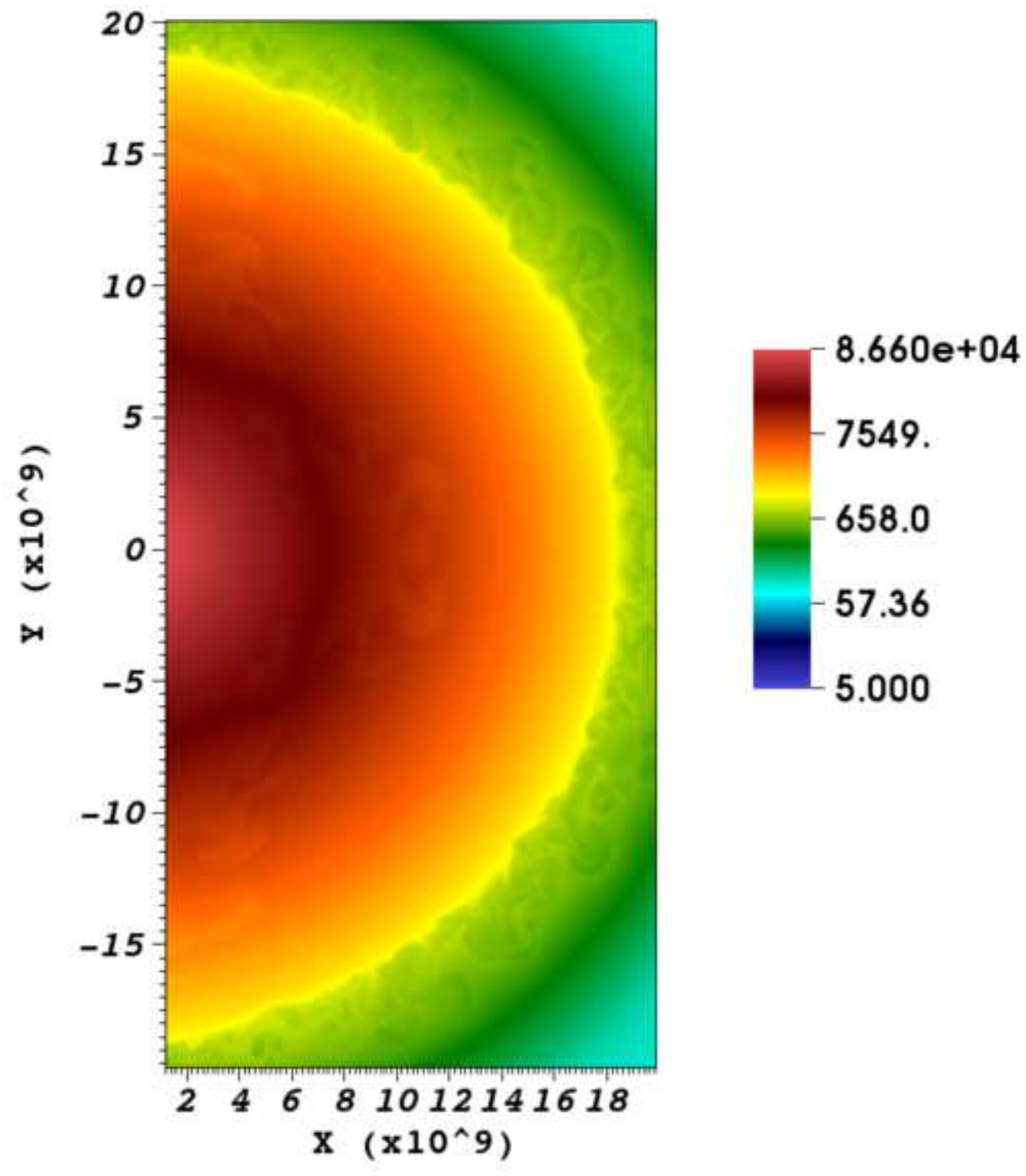,angle=0,width=4in}
}
\caption{Same as Figure 8 but for PISN model 195sm\_rotnoST ($Z =$~$10^{-4}$~$Z_{\odot}$) at time $t =$~ 80~s after 
the onset of dynamical collapse.}
\end{figure}

\begin{figure}       
\centerline{
\hskip -0.2 in
\psfig{figure=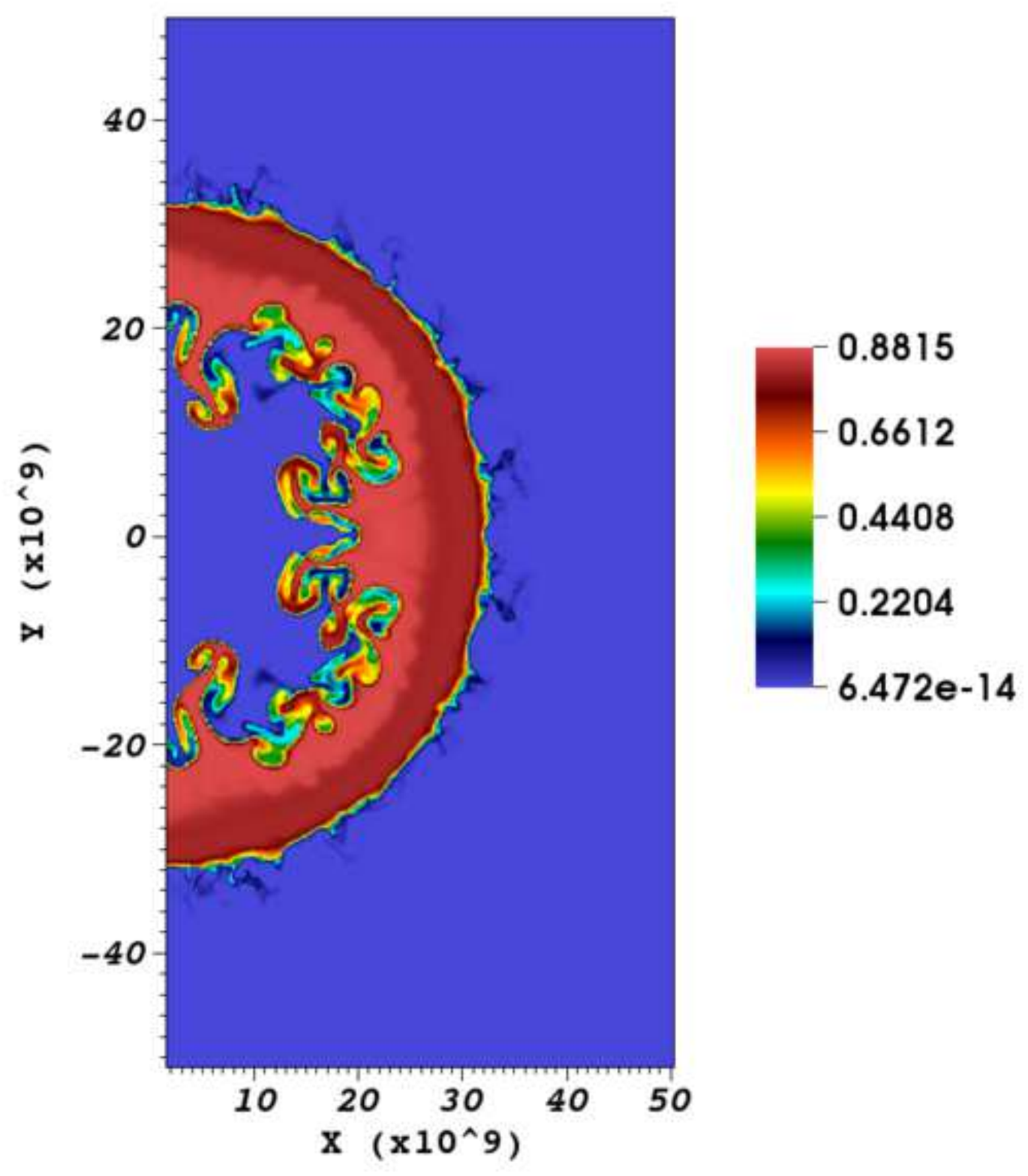,angle=0,width=4in}
\psfig{figure=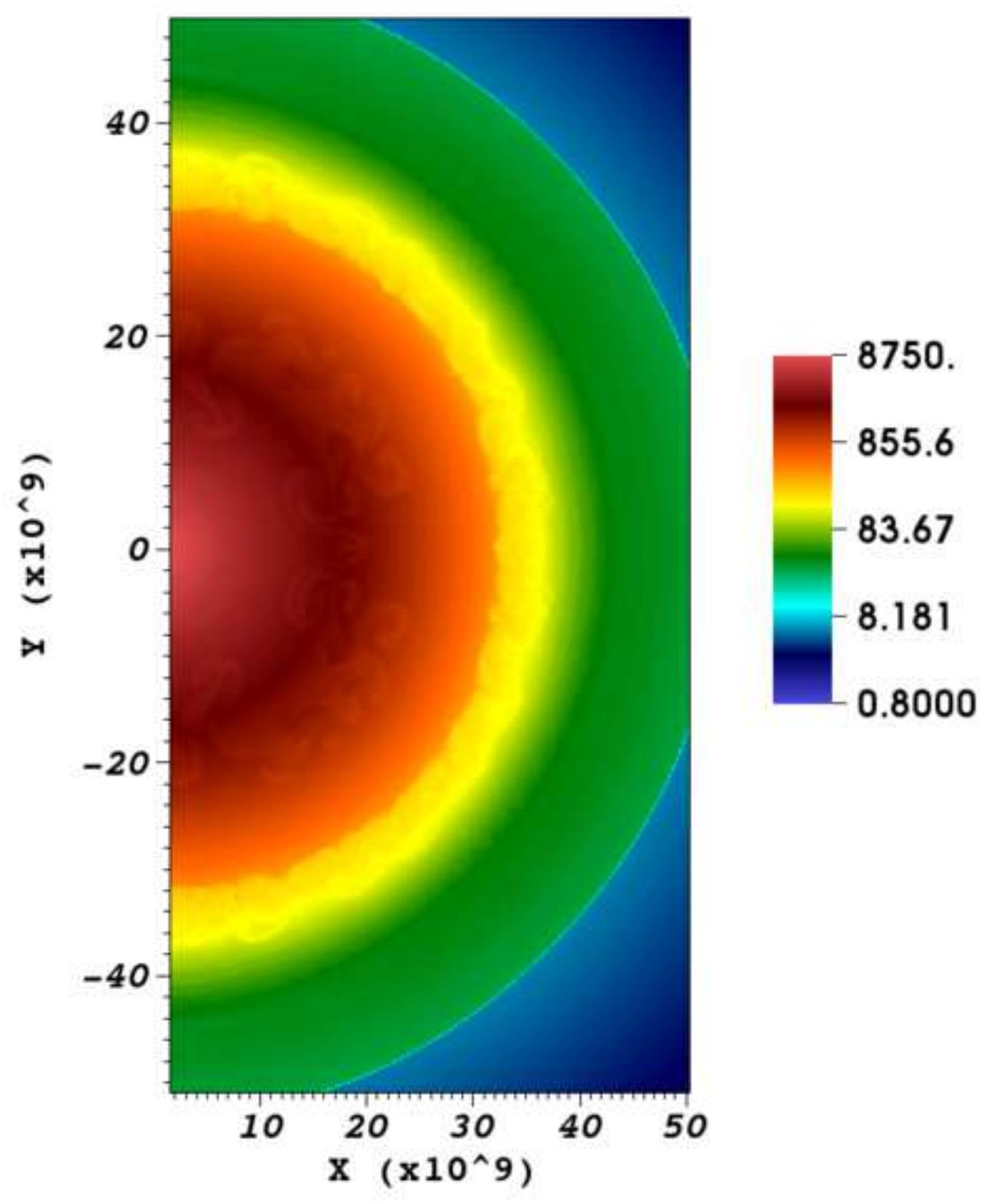,angle=0,width=4in}
}
\caption{Same as Figure 8 but for PISN model 217sm\_rotST\_ml2 ($Z =$~$10^{-4}$~$Z_{\odot}$) at time $t =$~ 105~s after 
the onset of dynamical collapse.}
\end{figure}

\begin{figure}
\begin{center}
\includegraphics[angle=0,width=19cm]{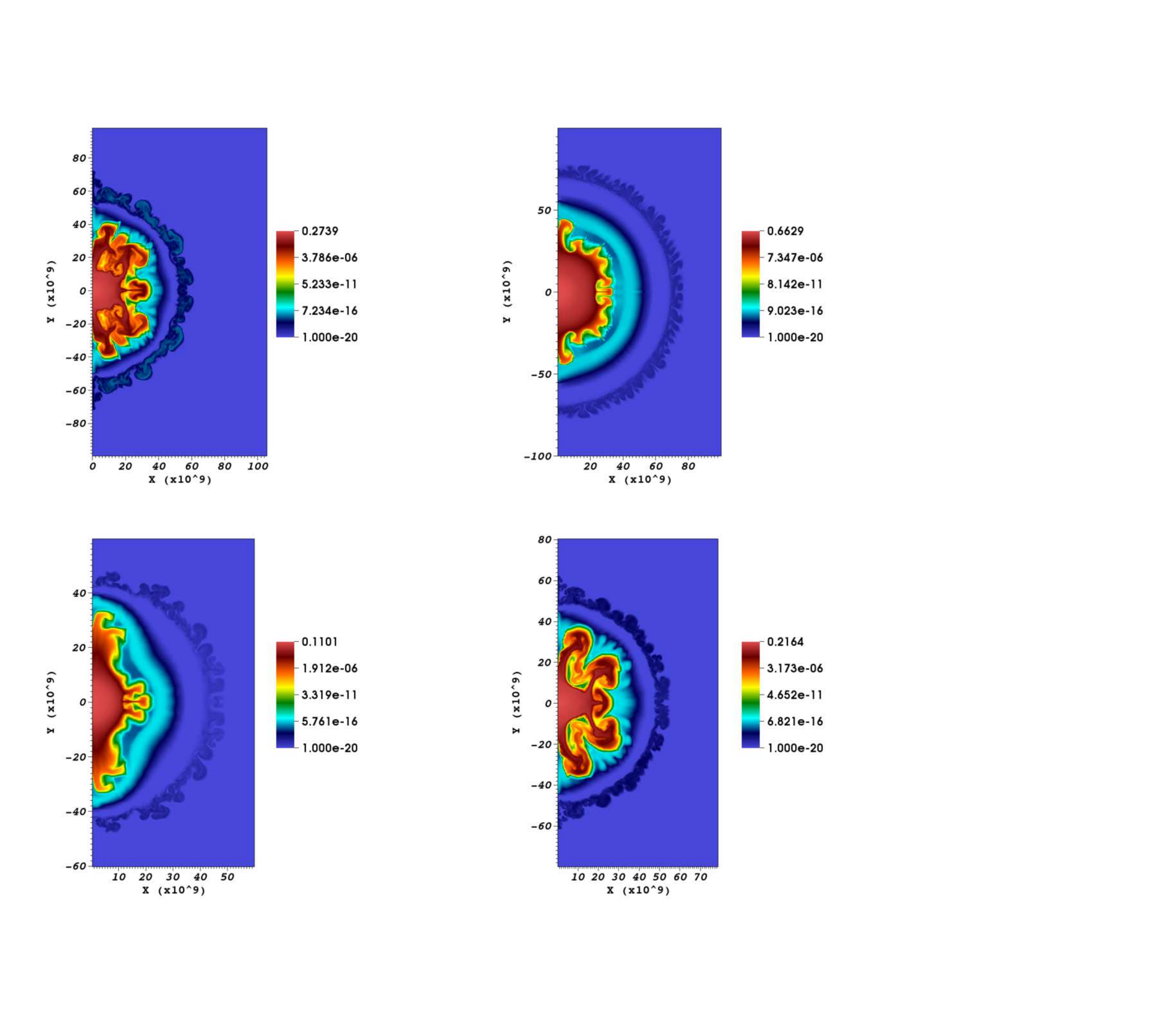}
\caption{$^{56}$Ni mass fraction for the $Z =$~$10^{-3}$~$Z_{\odot}$ PISN model series
at $\sim$~150~s after the onset of dynamical collapse: 200sm\_norot (upper left panel), 140sm\_rotST (upper right panel),
135sm\_rotnoST (lower left panel) and 150sm\_rotST\_ml2 (lower right panel).}
\end{center}
\end{figure}

\begin{figure}
\begin{center}
\includegraphics[angle=0,width=19cm]{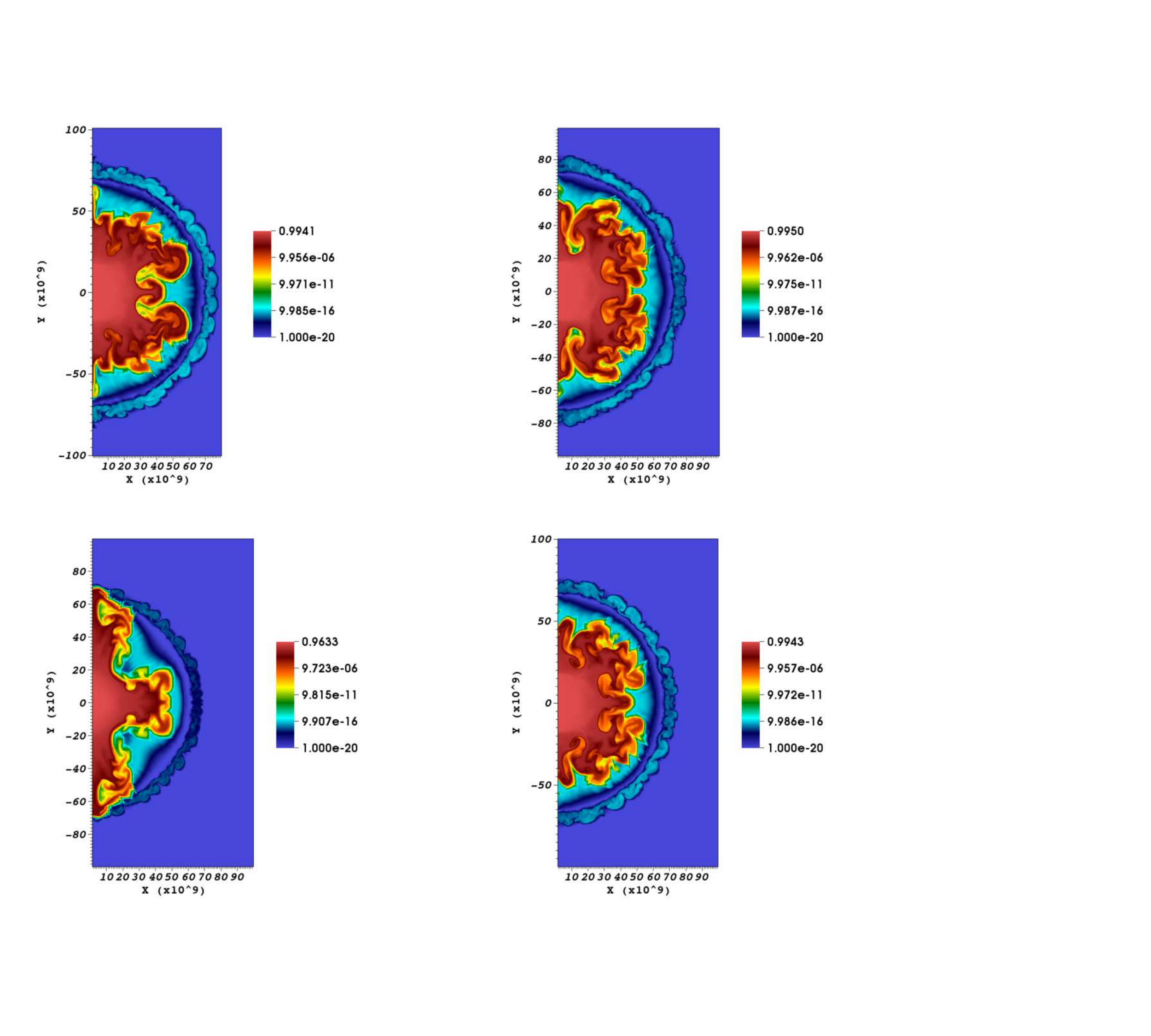}
\caption{Same as Figure 16 but for the $Z =$~$10^{-4}$~$Z_{\odot}$ PISN model series:
245sm\_norot (upper left panel), 205sm\_rotST (upper right panel),
195sm\_rotnoST (lower left panel) and 217sm\_rotST\_ml2 (lower right panel).}
\end{center}
\end{figure}

\begin{figure}
\begin{center}
\includegraphics[angle=0,width=19cm]{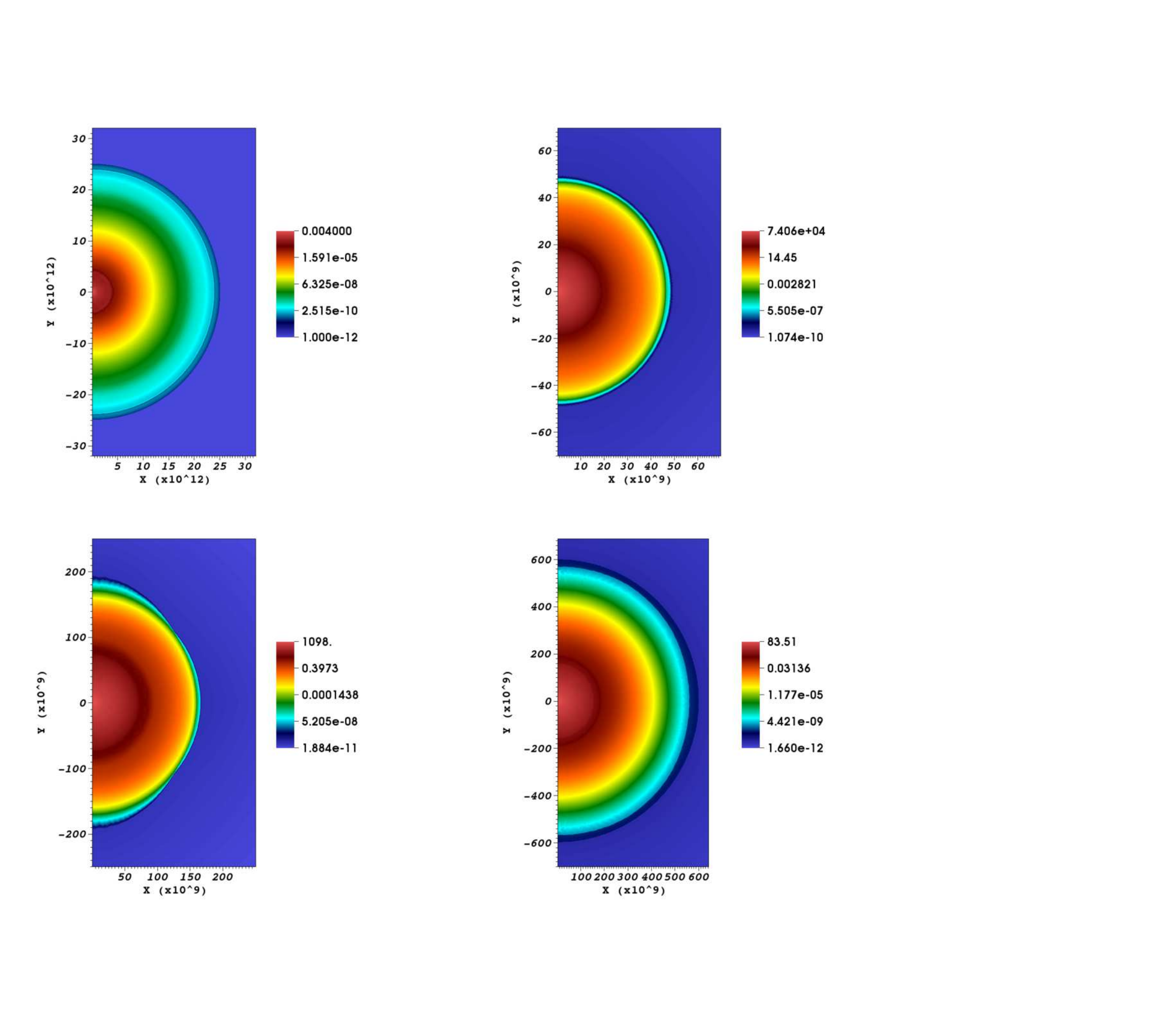}
\caption{Density of the $Z =$~$10^{-3}$~$Z_{\odot}$ PISN model series at
the times that the SN shock waves break out of the progenitor envelopes.}
\end{center}
\end{figure}

\begin{figure}
\begin{center}
\includegraphics[angle=0,width=19cm]{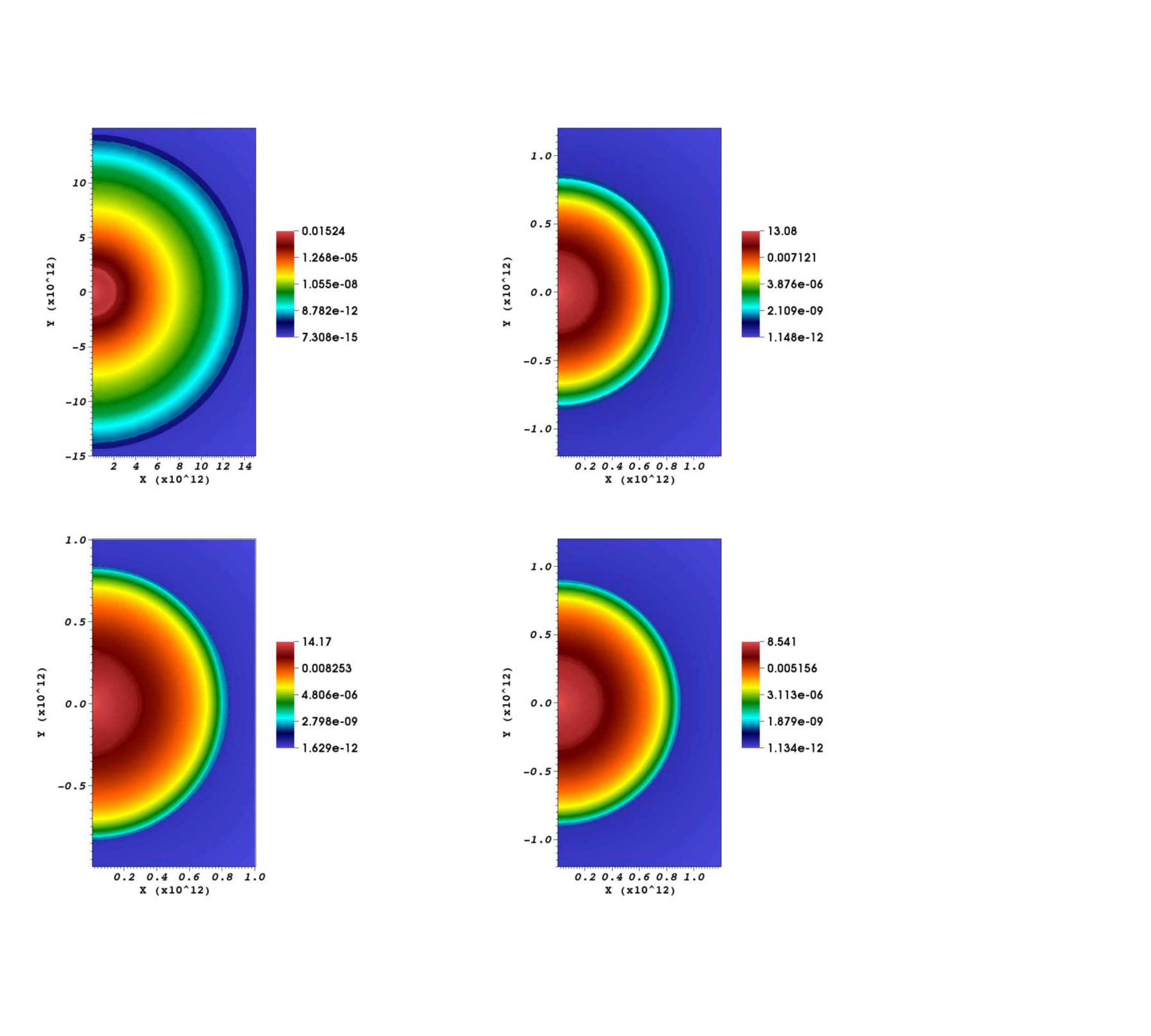}
\caption{Same as Figure 18 but for the $Z =$~$10^{-4}$~$Z_{\odot}$ PISN model series.}
\end{center}
\end{figure}

\begin{figure}
\begin{center}
\includegraphics[angle=0,width=16cm]{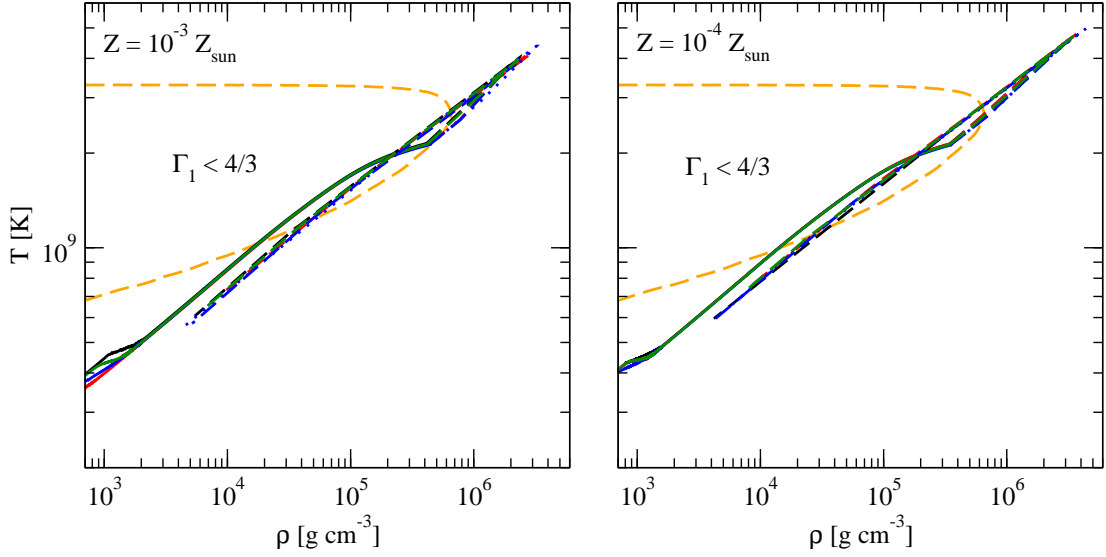}
\caption{Density-temperature structures of the $Z =$~$10^{-3}$~$Z_{\odot}$ (left panel) 
and $Z =$~$10^{-4}$~$Z_{\odot}$ (right panel) PISN model series. 
The solid curves represent the $\rho$-$T$ structure of the models at the time they encounter the dynamical pair-instability and 
are mapped into the 2-D {\it FLASH} AMR grid. The dashed curves show the subsequent dynamical 
evolution of $\rho_{c}$ and $T_{c}$. 
The thick orange dashed curve shows the area of $\Gamma_{1} <$~4/3 due to e$^{+}$e$^{-}$ pairs.
As with Figures 1-6, black curves represent ``norot", red curves the ``rotST", 
blue curves the ``rotnoST" and green curves the ``rotST\_ml2" models. The dotted blue curves represent
the rotating models without the effects of ST included, but with their rotational velocities artificially set to zero
upon mapping to {\it FLASH} (``rotnoST\_v0" models).}
\end{center}
\end{figure}

\begin{figure}
\begin{center}
\includegraphics[angle=0,width=16cm]{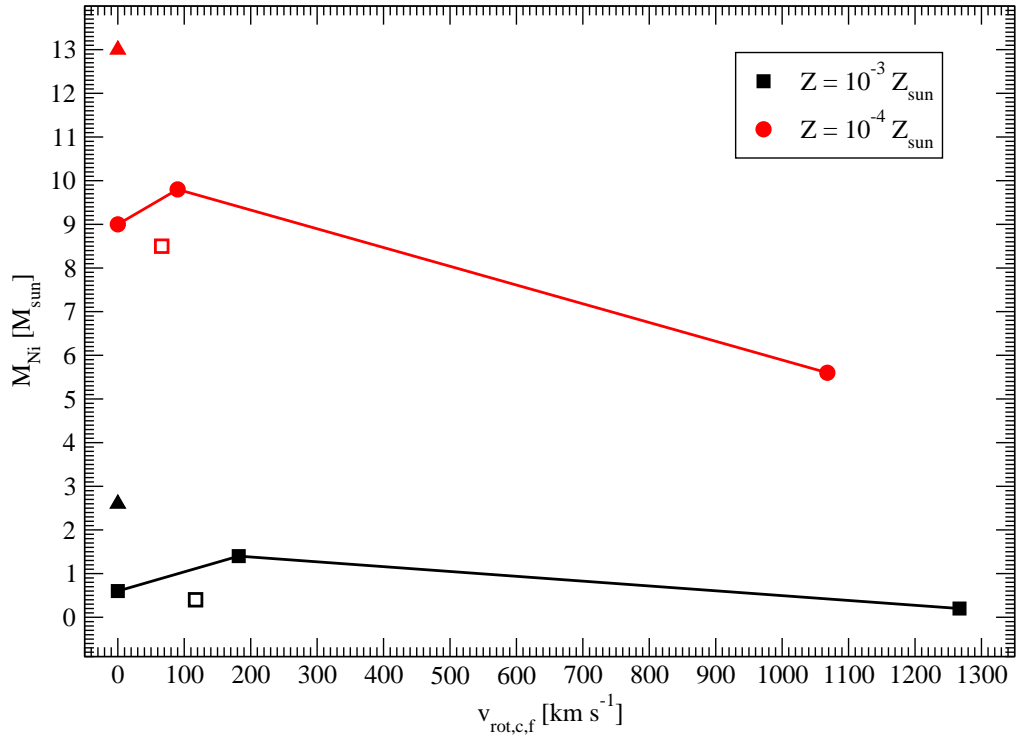}
\caption{Variation of the final PISN $^{56}$Ni mass with C/O core rotational 
velocity at time of mapping to {\it FLASH} as determined by our set of simulations. 
The filled triangles show the results for the ``rotnoST\_v0" models. The open squares
show the results for the ``rotST\_ml2" models.}
\end{center}
\end{figure}

\begin{figure}
\begin{center}
\includegraphics[angle=0,width=16cm]{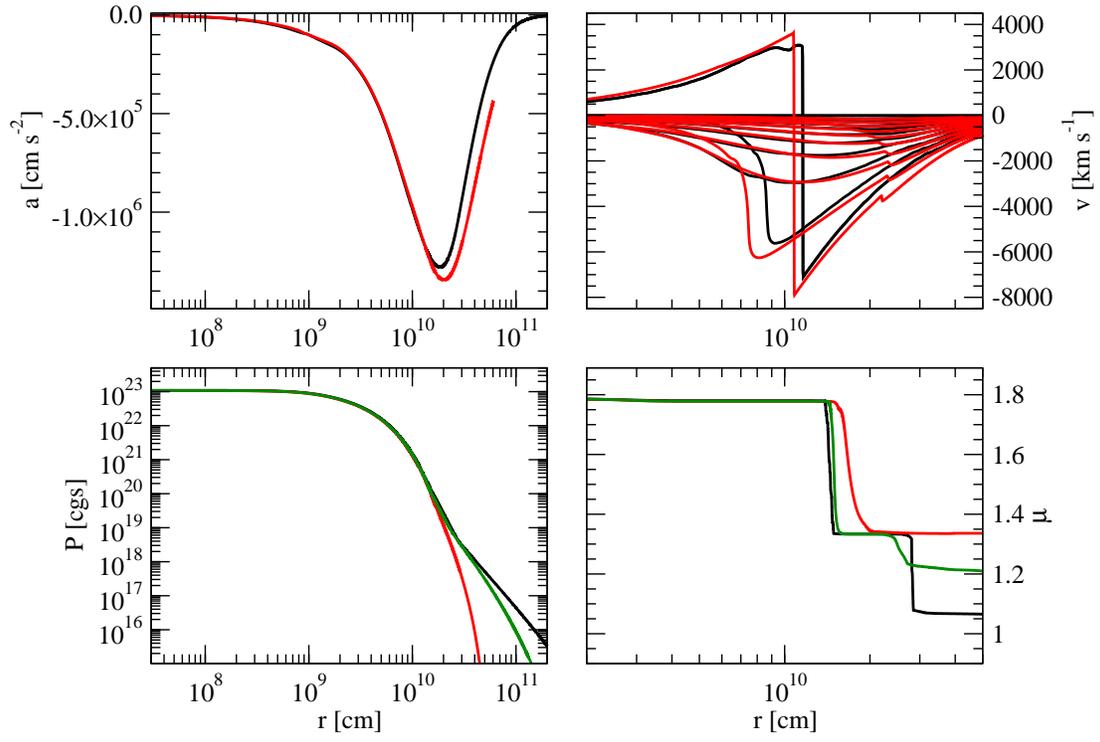}
\caption{Comparison of the dynamical collapse of the 200sm\_norot and the 140sm\_rotST models. 
Acceleration (upper left panel), radial velocity (upper right panel), pressure (lower left panel) and 
mean molecular weight $\mu$ (lower right panel) as a function of radius. The radial velocity is plotted
for 0-100~s in steps of ~10~s. Black curves correspond to the ``norot," red curves
to the ``rotST" and green curves to the ``rotST\_ml2" model.}
\end{center}
\end{figure}

\begin{figure}
\begin{center}
\includegraphics[angle=0,width=16cm]{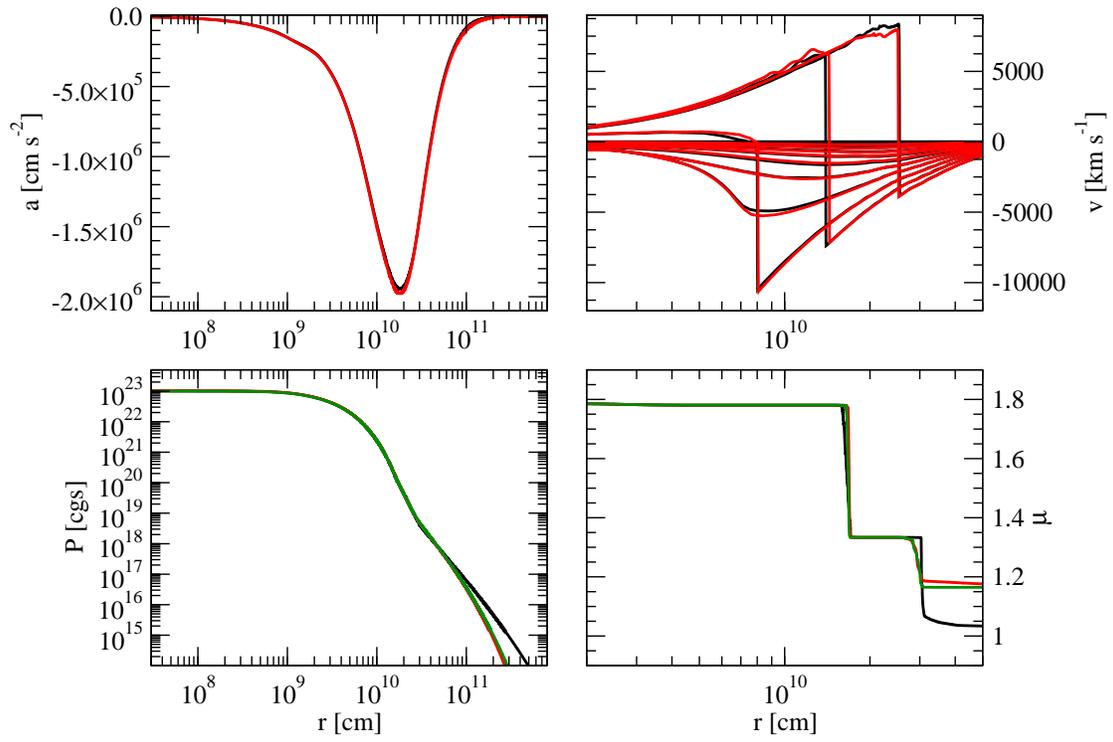}
\caption{Same as Figure 22 but for the 245sm\_norot and the 205sm\_rotST models. }
\end{center}
\end{figure}

\clearpage
%
\setcounter{table}{0}
\begin{deluxetable}{lccccccccc}
\tabletypesize{\tiny\tiny\tiny}
\tablewidth{0pt}
\tablecaption{Characteristics of PISN progenitor models.}
\tablehead{
\colhead {Model} &
\colhead {$M_{\rm ZAMS}$} &
\colhead {$M_{f}$} &
\colhead {$\Omega/\Omega_{c,s,i}$} &
\colhead {$\Omega/\Omega_{c,s,f}$} &
\colhead {$\Omega/\Omega^{a}_{c,c,f}$} &
\colhead {$v^{a}_{rot,c}$~(km~s$^{-1}$)} &
\colhead {$-E_{b,f}$~($10^{53}$~erg)} &
\colhead {$R_{f}$~($10^{11}$~cm)} 
 &\\}
\startdata
&&&&&$Z =$~$10^{-3}$~$Z_{\odot}$&&\\
\hline 
200sm\_norot			&  200.0   &  120.7   & 0.00 & 0.00 & 0.00 & 0.0     & 1.27 &  249.92 \\
140sm\_rotST			&  140.0   &   83.5  & 0.50 & 0.79 & 0.05 & 182.3   & 1.11 &  0.56   \\
135sm\_rotnoST			&  135.0   &   87.6  & 0.50 & 1.00 & 0.30 & 1266.5  & 1.14 &  1.66   \\
135sm\_rotnoST\_v0$^{\dagger}$		&  135.0   &   87.6  & 0.50 & 0.00 & 0.00 & 0.0     & 1.14 &  1.66   \\
150sm\_rotST\_ml2		&  150.0   &   93.1  & 0.50 & 0.99 & 0.03 & 117.4   & 1.18 &  5.99   \\
\hline
&&&&&$Z =$~$10^{-4}$~$Z_{\odot}$&&\\
\hline
245sm\_norot			&  245.0   &  141.8  & 0.00 & 0.00 & 0.00 & 0.00    & 1.71 &  104.01 \\
205sm\_rotST			&  205.0   &  123.0  & 0.50 & 0.99 & 0.02 & 90.00   &1.71  &  8.61   \\
195sm\_rotnoST			&  195.0   &  121.5    & 0.50 & 0.13 & 0.25 & 1067.80 & 1.66 &  8.39   \\
195sm\_rotnoST\_v0$^{\dagger}$		&  195.0   &  121.5  & 0.50 & 0.00 & 0.00 & 0.00    & 1.66 &  8.39   \\
217sm\_rotST\_ml2		&  217.0   &  122.6  & 0.50 & 0.56 & 0.02 & 65.60   & 1.67 &  9.03   \\
\enddata 
\tablecomments{All masses are expressed in $M_{\odot}$. 
$^{a}$ We define as the ``edgeÓ of the CO core the radius at which $X_{C} + X_{O} <$~0.5. 
$^{\dagger}$ The rotational velocities of these pre-PISN models were artificially set to zero for the 
{\it FLASH} hydro simulations in order to investigate the effects of rotation in otherwise identical models.}
\end{deluxetable}

\setcounter{table}{1}
\begin{deluxetable}{lccccccccccc}
\tabletypesize{\tiny\tiny\tiny}
\tablewidth{0pt}
\tablecaption{Characteristics of PISN explosions.}
\tablehead{
\colhead {Model} &
\colhead {$^{4}$He} &
\colhead {$^{12}$C} &
\colhead {$^{16}$O} &
\colhead {$^{20}$Ne} &
\colhead {$^{24}$Mg} &
\colhead {$^{28}$Si} &
\colhead {$^{32}$S} &
\colhead {$\rho^{a}_{c,max,6}$} &
\colhead {$T^{b}_{c,max,9}$} &
\colhead {$^{56}$Ni$_{f}$} 
 &\\}
\startdata
&&&&&$Z =$~$10^{-3}$~$Z_{\odot}$&&&&\\
\hline 
200sm\_norot				&  35.5 & 3.7 & 65.5 & 8.0  & 2.3 & 0.1 & $<$0.0001 & 2.353 & 4.024 & 0.6 \\
140sm\_rotST				&  3.8  & 5.0 & 66.3 & 6.7  & 1.6 & 0.1 & $<$0.0001 & 2.879 & 4.126 & 1.4 \\
135sm\_rotnoST				&  7.5  & 4.4 & 66.4 & 7.2  & 1.8 & 0.1 & $<$0.0001 & 2.183 & 3.837 & 0.2 \\
135sm\_rotnoST\_v0			&  7.5  & 4.4 & 66.4 & 7.2  & 1.8 & 0.1 & $<$0.0001 & 3.229 & 4.377 & 2.6 \\
150sm\_rotST\_ml2			&  14.1 & 4.4 & 65.4 & 6.9  & 1.7 & 0.1 & $<$0.0001 & 2.309 & 3.971 & 0.4 \\
\hline	
&&&&&$Z =$~$10^{-4}$~$Z_{\odot}$&&&&\\
\hline
245sm\_norot				&  37.2 & 4.4 & 80.8 & 10.2 & 3.1 & 0.1 & $<$0.0001 & 3.546 & 4.727 & 9.0 \\
205sm\_rotST				&  21.7 & 4.5 & 82.1 & 10.3 & 3.1 & 0.1 & $<$0.0001 & 3.670 & 4.777 & 9.8 \\
195sm\_rotnoST				&  20.5 & 4.3 & 81.8 & 10.5 & 3.2 & 0.1 & $<$0.0001 & 3.132 & 4.525 & 5.6 \\
195sm\_rotnoST\_v0			&  20.5 & 4.3 & 81.8 & 10.5 & 3.2 & 0.1 & $<$0.0001 & 4.378 & 4.968 &13.0 \\
217sm\_rotST\_ml2			&  22.8 & 4.2 & 80.6 & 10.3 & 3.2 & 0.1 & $<$0.0001 & 3.529 & 4.714 & 8.5 \\
\enddata 
\tablecomments{All species masses are in units of $M_{\odot}$ and correspond to the total mass included in the simulation
box.
$^{a}$ In units of $10^{6}$~g~cm$^{-3}$.
$^{b}$ In units of $10^{9}$~K.}
\end{deluxetable}


\end{document}